\documentclass[reprint, amsmath, amssymb, aps, prapplied, superscriptaddress]{revtex4-2}

\usepackage{graphicx} 
\usepackage{dcolumn}  
\usepackage{bm}      
\usepackage[hidelinks]{hyperref} 
\usepackage[mathlines]{lineno} 
\usepackage{placeins} 
\usepackage{braket}

\usepackage{tikz}
\usetikzlibrary{arrows.meta,positioning,calc}

\newlength{\tikzbbpad}
\tikzset{
  bbpad/.style={
    execute at end picture={
      \path (current bounding box.south west) ++(-\tikzbbpad,-\tikzbbpad) coordinate (bbSW);
      \path (current bounding box.north east) ++(\tikzbbpad,\tikzbbpad) coordinate (bbNE);
      \useasboundingbox (bbSW) rectangle (bbNE);
    }
  }
}

\begin{document}

\title{Beyond Single-Shot Fidelity: Chernoff-Based Throughput Optimization in Superconducting Qubit Readout}

\author{Sinan Bugu}
	\email{sbugu@ncsu.edu}
	\email{sinanbugu@gmail.com}
\affiliation{Department of Physics and Astronomy, North Carolina State University, Raleigh, North Carolina, USA}
\affiliation{BuQuLab Research Laboratory, Winston-Salem, North Carolina, USA}

\date{\today}

\begin{abstract}
Single-shot fidelity is the standard benchmark for superconducting qubit readout, but it does not directly minimize the total wall-clock time required to certify a qubit state to a specified target error. We develop a theoretical and numerical description of dispersive readout by treating the measurement record as a stochastic communication channel. Within a model of dispersive readout that includes $T_1$ relaxation and the finite cavity response time, we compute the classical Chernoff information governing the multi-shot error exponent. We find a separation between the integration time that maximizes single-shot fidelity and the time that maximizes the asymptotic Chernoff rate, whose magnitude and ordering depend on the per-shot overhead. Across the main surveys, which use per-shot overheads of $\tau_{\mathrm{oh}}=5$--$30~\mu\mathrm{s}$, the rate-optimal integration window $\tau_{C}$ is longer than the fidelity-optimal one, with parameter-dependent asymptotic rate gains of approximately $12.8$--$15.7\%$ at standard readout linewidth ($\kappa/2\pi=5$~MHz) and $6.3$--$11.6\%$ across the corresponding sub-MHz-linewidth survey. For a specified target error $\epsilon$, however, certification requires an integer number of shots. The exact certification time therefore exhibits integer-shot threshold structure. Its target-specific optimum lies at or below $\tau_{C}$ and converges toward $\tau_{C}$ asymptotically, although its drift over practical targets is not strictly monotonic. We therefore adopt a Bahadur--Rao-based preset, selected at the shallowest candidate design target within a declared target interval, which lowers both the overall median and worst-band median calibration regret relative to $\tau_{C}$ at every surveyed parameter point. Comparing the extracted classical information to the unit-efficiency Gaussian Chernoff benchmark defines an information-extraction efficiency metric. For the baseline detection efficiency $\eta=0.45$, the information-extraction efficiency approaches $45\%$ at short integration times and decreases to $\eta_{\mathrm{info}}(\tau_{C})\approx 10.8\%$ at $\tau_{C}\approx 0.263~\mu\mathrm{s}$ because of $T_1$-induced trajectory smearing. This formulation connects readout calibration to the operational objective of minimizing certification time in high-throughput superconducting processors.
\end{abstract}

\keywords{Superconducting qubits, Dispersive readout, Chernoff error exponent, Quantum hypothesis testing, Readout calibration, Throughput optimization, Finite-shot calibration, Bahadur--Rao preset, Transmon}

\maketitle

\section{Introduction}

The scaling of superconducting quantum processors toward the fault-tolerant regime requires not only high-fidelity operations but also high-throughput certification cycles \cite{fowler2012surface, terhal2015quantum}. Recent surface-code demonstrations have shown that cycle time is a critical scaling bottleneck \cite{Acharya2025, krinner2022realizing}, with system-level latencies in large-scale processors often dominating total execution time \cite{Arute2019}. Within the circuit quantum electrodynamics (cQED) architecture, dispersive readout of transmon qubits remains one of the most persistent performance constraints \cite{blais2021circuit, walter2017rapid}, despite advances in dynamic control of the dispersive interaction that push readout speed and fidelity \cite{Swiadek2024}. Although coherent gates now operate on nanosecond timescales, measurement windows remain disproportionately long relative to intrinsic device dynamics \cite{clerk2010introduction, gambetta2008quantum}.

This latency reflects a trade-off inherent to dispersive measurement. Increasing the measurement drive to accelerate information extraction enhances measurement-induced dephasing, AC Stark shifts, and transitions into non-computational leakage manifolds \cite{reed2010high, murch2013observing, hatridge2013quantum, fechant2025offset}. At sufficiently high drive powers, these effects can escalate into nonlinear phenomena such as transmon ionization \cite{Dumas2024} and stochastic state transitions arising from breakdown of the dispersive approximation \cite{nesterov2024measurement}. Alternative coupling architectures have therefore been explored to enable faster measurement while mitigating backaction, including longitudinal interaction schemes \cite{didier2015fast} and the quarton coupler enabling nanosecond-scale readout windows \cite{Ye2024_Quarton}. Complementary mitigation strategies, such as integrated filtering and Purcell-protected device designs, further suppress measurement-induced decay and enhance stability in high-power regimes \cite{Sah2024}.

In standard practice, readout is optimized by maximizing single-shot Bayes fidelity \cite{jeffrey2014fast}. More recently, model-based approaches have enabled systematic suppression of assignment and leakage errors in multi-qubit architectures \cite{Bengtsson2024}. However, for protocols involving repeated measurements, single-shot fidelity does not fully capture the cumulative effects of leakage and hardware latency \cite{Hazra2025}. While multiplexed readout \cite{heinsoo2018rapid, Spring2025}, active reset \cite{bultink2016active}, and near-quantum-limited amplification \cite{castellanos2008amplification, macklin2015near} have improved raw measurement performance, scalable architectures ultimately seek to minimize the wall-clock time required to reach a target confidence level $\epsilon$. For repeated-measurement protocols, the relevant objective shifts from isolated decision accuracy to certification throughput. Consequently, the fidelity-optimal integration time $\tau_{\text{fid}}$ may be suboptimal once hardware latency and reset constraints are incorporated \cite{riste2012feedback}.

We analyze readout using the full measurement-score distributions rather than isolated binary outcomes. Chernoff information as a figure of merit for analog qubit readout was established by D'Anjou \cite{danjou2021generalized}, who showed that it generalizes assignment fidelity to non-Gaussian statistics and to repeated measurements, and its asymptotic role follows from hypothesis-testing theory \cite{Chernoff1952, nussbaum2009asymptotic}. We do not claim to introduce Chernoff information as a readout metric. Our contribution is threefold. First, we treat $C$ as an explicit function of the integration time within a continuous-time dispersive model that includes $T_1$ relaxation and the finite cavity response time, rather than as a single number characterizing a fixed measurement. Second, we optimize $C(\tau)$ per unit wall-clock time by including a per-shot hardware overhead, which converts the readout figure of merit into a throughput objective and relocates the optimum. Third, we replace the asymptotic exponent by exact $N$-shot error probabilities, so that the integer-shot structure of certification at a declared target error is treated explicitly rather than through the error exponent alone. Chernoff information has also been used to optimize the integration time in the duty cycle of nitrogen-vacancy-center readout \cite{panadero2023photon}, without quantifying the shift relative to the single-shot-fidelity optimum or the role of a per-shot overhead.

Our central contribution is to formulate readout calibration as a wall-clock certification problem. Throughout, ``certification'' denotes repeated binary discrimination between independently prepared ground- and excited-state hypotheses, and $\epsilon$ denotes the target error probability of that decision. For $N$ independent shots the Chernoff bound gives $\epsilon \sim e^{-C(\tau)N}$, so reaching a target $\epsilon$ requires $N = \log(1/\epsilon)/C(\tau)$ shots, where $C(\tau)$ is the classical Chernoff information. Each shot occupies $\tau+\tau_{\text{oh}}$ of wall-clock time, with $\tau_{\text{oh}}$ the fixed per-shot overhead, so the total time $N(\tau+\tau_{\text{oh}})$ is the asymptotic certification-time proxy
\begin{equation}
\bar{T}_{\text{cert}}(\tau) = \frac{\log(1/\epsilon)}{C(\tau)}(\tau+\tau_{\text{oh}}),
\end{equation}
whose minimizer is independent of $\epsilon$. Rather than maximizing single-shot fidelity, we first minimize this proxy. Because certification at a specified target consumes an integer number of shots, we then treat the exact finite-shot objective separately and quantify the \emph{calibration regret}, the fractional excess certification time, incurred by fixed integration-time rules. Consistent with experimental observations \cite{jeffrey2014fast, walter2017rapid}, we model homodyne score distributions using Gaussian noise kernels, subsequently deformed by $T_1$ relaxation events that produce asymmetric tails and assignment errors \cite{gambetta2008quantum}. Using a model of dispersive readout that includes $T_1$ relaxation and the finite cavity response time \cite{boutin2017resonator}, we demonstrate a systematic separation between the fidelity-optimal integration time $\tau_{\text{fid}}$ and the rate-optimal time $\tau_{C}$, a separation that persists deep into the long-coherence regime.

We also derive asymptotic scaling laws showing that any nonzero hardware overhead shifts the rate optimum toward longer integration times. This shift amortizes the fixed temporal cost of each shot. We propose a diagnostic, evaluable from measured score distributions, that compares the extracted Chernoff information with the unit-efficiency Gaussian benchmark \cite{audenaert2007discriminating, pirandola2008computable}. The resulting calibration procedure applies to high-duty-cycle superconducting processors, where power-induced deviations from the dispersive regime \cite{sank2016measurement} require careful optimization of temporal resources. Photon-noise-tolerant readout designs address the same trade-off between speed, noise, and nonlinearity \cite{Sunada2024}. For orientation we use the following terms consistently: a \emph{preset} is a single fixed integration time adopted before the target error is known; the \emph{survey} is the set of twenty parameter points over which all calibration rules are compared; a \emph{baseline} is the single parameter point used for the illustrative figures; and \emph{standard-linewidth} and \emph{sub-MHz-linewidth} label the two cavity-parameter regimes specified at the start of Sec.~III.

\section{Stochastic Communication Model of Dispersive Readout}

\subsection{Dispersive Dynamics and Homodyne SNR}

We model the dispersive readout of a transmon qubit coupled to a readout resonator, described by the Hamiltonian $H/\hbar = \omega_r a^\dagger a + \frac{\omega_q}{2} \sigma_z + \chi a^\dagger a \sigma_z$ \cite{blais2020quantum}. Under a constant microwave drive, we represent the two qubit-conditioned measured-quadrature responses by the normalized branches $\pm\Delta\alpha(t)$. The branch-response function is
\begin{equation}
    \Delta\alpha(t) = \frac{4\chi}{\kappa} \left( 1 - e^{-\kappa t / 2} \right)
    \label{eq:branch}
\end{equation}
where $\chi$ is the dispersive shift and $\kappa$ is the cavity linewidth \cite{gambetta2008quantum}. Here $\Delta\alpha(t)$ is the dimensionless branch-response function used in the normalized score model, rather than an absolute coherent-state amplitude: it is the scalar response that enters the matched filter $h(t)\propto\Delta\alpha(t)$ and the SNR of Eq.~\eqref{eq:snr}, and the $\pm$ notation is the symmetric score-coordinate convention of the reduced model. Any alternative response model is normalized to enter these same definitions in the same convention. The separation of the two branches in these units is $2\Delta\alpha(t)$, while the physical signal scale proportional to $\sqrt{\bar n}$ is carried through $\bar n$ in the SNR definition below. Because homodyne SNR conventions depend on the normalization of both the measured quadrature and the score variable, the following equation defines the convention used consistently throughout this work. Equation~\eqref{eq:branch} is a reduced single-pole response model. For the abrupt drive of an initially empty cavity assumed below, the exact dispersive response has a quadratic rather than a linear onset and a steady state smaller by $[1+(2\chi/\kappa)^{2}]^{-1/2}$; Appendix~\ref{app:response} derives it, states the approximations involved, and shows that the ordering and qualitative scale of the separation between the two optima are preserved. The information-theoretic framework itself is independent of this reduced response model: the score distributions entering the Chernoff and finite-shot calculations may instead be supplied by a more complete dynamical model or directly by calibrated experimental I/Q data.

The information reaching the homodyne detector is processed via a matched filter to maximize the signal-to-noise ratio (SNR) \cite{clerk2010introduction, boutin2017resonator}. The filter weighting is chosen to match the branch-response shape, $h(t)\propto\Delta\alpha(t)$; any common scale factor cancels from the trajectory projection normalized to the no-decay case. For a given integration time $\tau$, the squared signal-to-noise ratio ($SNR^2$) at the detector is defined as
\begin{equation}
    SNR^2(\tau)
    =
    \eta \kappa \bar{n}
    \int_0^\tau |\Delta\alpha(t)|^2 dt .
    \label{eq:snr}
\end{equation}
where $\eta$ is the overall quantum efficiency and $\bar{n}$ is the steady-state photon number. In the present model, the single parameter $\tau$ therefore fixes three intervals that are independently settable in experiment: the duration of the readout drive, the opening of the acquisition window, and the extent of the matched-filter record. We assume a constant drive that begins at $t=0$, acquisition that begins with the drive, and a matched filter applied to the full record on $[0,\tau]$. No high-amplitude ring-up segment, acquisition delay, or truncated integration sub-window is modelled. This is the appropriate description when the pulse envelope and the acquisition onset are held fixed by prior calibration and only the terminal integration cutoff is varied, which is the one-dimensional calibration slice studied here. The matched filter already down-weights the early record because $h(t)\propto\Delta\alpha(t)$ vanishes at $t=0$. A general treatment would replace $\tau$ by a timing vector $\boldsymbol{\theta}=(t_{\mathrm{ring}},t_{\mathrm{delay}},t_{\mathrm{int}})$ and maximize $\mathcal{R}_C(\boldsymbol{\theta})=C(\boldsymbol{\theta})/ T_{\mathrm{cycle}}(\boldsymbol{\theta})$. The wall-clock objective of Eq.~\eqref{eq:t_cert} and the finite-shot construction of Sec.~II.D carry over unchanged; only the search space grows. We define the matched-filter score as a normalized projection variable whose conditional noise variance is unity. In this convention, detector inefficiency $\eta$ enters exclusively through the mean separation encoded in $\mathrm{SNR}(\tau)$, while the additive noise variance of the score distribution is fixed to one. While we assume a standard homodyne detection chain, alternative high-efficiency detection schemes using on-chip microwave photon counters have demonstrated comparable performance by mapping quantum information directly into binary click statistics \cite{opremcak2021high}.

\subsection{Stochastic Trajectory Smearing with Finite Cavity Response}

To accurately represent the excited-state distribution $p_e(x)$, we account for stochastic relaxation of the qubit from $|e\rangle$ to $|g\rangle$ at a jump time $t_j\in[0,\tau]$. We adopt a trajectory model that ensures continuity of the measured-quadrature branch coordinate at the jump time. For $t<t_j$, the system follows the excited-state trajectory $\alpha(t)=+\Delta\alpha(t)$. At the jump, the branch coordinate is $+\Delta\alpha(t_j)$. For $t>t_j$, it relaxes exponentially toward the ground-state trajectory $-\Delta\alpha(t)$ at a rate governed by $\kappa/2$:
\begin{equation}
    \alpha(t > t_j) = -\Delta\alpha(t) + 2\Delta\alpha(t_j) e^{-\kappa(t-t_j)/2}
\end{equation}
This expression guarantees continuity of the quadrature trajectory at the jump boundary, $\alpha(t_j^-)=\alpha(t_j^+)=\Delta\alpha(t_j)$. The cavity response time is what makes the post-jump record informative: the intracavity field does not follow the qubit instantaneously, so after a decay event the measured quadrature relaxes toward the ground-state branch at the rate $\kappa/2$ rather than switching abruptly. We assume that the readout drive remains on throughout the integration window, so after a decay event the cavity relaxes toward the driven ground-state trajectory rather than the undriven vacuum. The normalized trajectory projection is obtained by projecting this continuous trajectory onto the filter $h(t)$; the corresponding conditional score mean entering Eq.~\eqref{eq:excited_score_distribution} is $\mu(t_j) = [\mathrm{SNR}(\tau)/2]\, s(t_j)$, where $s(t_j)$ denotes the projection normalized to the no-decay case. In this unit-variance score convention, the no-decay conditional means are
\begin{equation}
\mu_e=\frac{\mathrm{SNR}(\tau)}{2},
\qquad
\mu_g=-\frac{\mathrm{SNR}(\tau)}{2},
\end{equation}
and the ground-state score distribution is $p_g(x)=\mathcal{N}(\mu_g,1)$. The full excited state distribution is constructed as:
\begin{equation}
    p_e(x) = e^{-\tau/T_1} \mathcal{N}(\mu_e,1)
    + \int_0^\tau \frac{1}{T_1} e^{-t_j/T_1}
    \mathcal{N}(\mu(t_j),1) dt_j
    \label{eq:excited_score_distribution}
\end{equation}
where the first term accounts for trajectories where no decay occurs during the integration window.

\subsection{Certification Wall-Clock Time Objective}

Related work on repeated quantum decision and coordination problems has considered performance measures beyond isolated single-shot accuracy \cite{bugu2025entanglement, bugu2026hfc}. In the present setting, we formulate certification as binary discrimination between independently prepared ground- and excited-state hypotheses, represented by $p_g(x)$ and $p_e(x)$, under equal prior probabilities. Here ``certification'' refers to this repeated binary decision problem rather than to general quantum-state verification, and successive shots are assumed to be independent after reset and re-preparation. Distinguishability is quantified using the classical Chernoff information \cite{Chernoff1952, CoverThomas2006}:
\begin{equation}
    C = -\log \left( \min_{s \in [0,1]} \int p_g(x)^s p_e(x)^{1-s} dx \right)
\end{equation}
Throughout this work, $\log$ denotes the natural logarithm, so that $C$ is expressed in nats; base-ten logarithms are written explicitly as $\log_{10}$.

A natural operational objective is the wall-clock time required to certify the qubit state to a target error probability $\epsilon$. Incorporating the hardware-specific overhead $\tau_{\text{oh}}$, which accounts for latencies and reset times \cite{Acharya2025, krinner2022realizing}, the asymptotic certification-time proxy is:
\begin{equation}
    \bar{T}_{\text{cert}}(\tau) = \frac{\log(1/\epsilon)}{C(\tau)} \left( \tau + \tau_{\text{oh}} \right)
      \label{eq:t_cert}
\end{equation}
Equation~\eqref{eq:t_cert} is asymptotic in two ways. First, $C(\tau)$ characterizes only the large-shot error exponent. Second, $\log(1/\epsilon)/C(\tau)$ is the continuous-valued shot count predicted from that exponent alone. Its minimizer is therefore exactly independent of $\epsilon$. We refer to this minimizer as the rate-optimal integration time
\begin{equation}
    \tau_{C} = \arg\max_{\tau}\, \mathcal{R}_C(\tau), \qquad
    \mathcal{R}_C(\tau) \equiv \frac{C(\tau)}{\tau+\tau_{\text{oh}}},
    \label{eq:tau_C}
\end{equation}
where $\mathcal{R}_C$ is the asymptotic Chernoff rate per unit wall-clock time. We stress that $\tau_{C}$ is defined by the asymptotic rate objective and need not minimize the exact certification time at a finite target; the exact objective is defined in Sec.~II.D. Comparing $\tau_{C}$ to the fidelity-optimal time $\tau_{\text{fid}}$, obtained by maximizing $F = 1 - \frac{1}{2} \int \min(p_g, p_e) dx$, defines the asymptotic rate-gain factor
\begin{equation}
    G = \frac{\bar{T}_{\text{cert}}(\tau_{\text{fid}})}{\bar{T}_{\text{cert}}(\tau_{C})}
      = \frac{\mathcal{R}_C(\tau_{C})}
       {\mathcal{R}_C(\tau_{\mathrm{fid}})}.
       \label{eq:gain}
\end{equation}
The corresponding percentage improvement is $100(G-1)\%$. The factor $G$ is exactly independent of $\epsilon$ and quantifies the asymptotic rate advantage over standard single-shot optimization in the presence of the per-shot overhead. This figure of merit presumes that the analog score $x$ is retained from shot to shot and combined at the level of the likelihood ratio. If each shot is instead reduced to a binary assignment before repeated-shot inference, the relevant per-shot quantity is the Chernoff information of the induced binary channel, computed from the two conditional assignment probabilities rather than from $p_g$ and $p_e$, and the optimum must be determined from that channel. The gains quantified here therefore pertain to architectures that retain analog or soft readout information across repeated shots. To isolate the physical degradation of the measurement channel, we define a unit-efficiency Gaussian Chernoff benchmark
\begin{equation}
C_{\mathrm{ideal}}(\tau) = \frac{1}{8} \kappa \bar{n} \int_0^\tau |\Delta\alpha(t)|^2 dt,
\end{equation}
which corresponds to the classical Chernoff information in the absence of $T_1$ relaxation and with detection efficiency $\eta = 1$. This benchmark represents the ideal distinguishability limit for a homodyne receiver observing a non-decaying qubit ($T_1 \to \infty$), and provides a baseline against which we quantify the information loss attributable to detector inefficiency and $T_1$-induced trajectory smearing. All reported curves are obtained from deterministic numerical integration of the score distributions without stochastic sampling. We verified convergence with respect to the score grid, the time grid, the decay-time quadrature, and the integration-time scan, refining both optima locally. Residual numerical errors are smaller than the marker size in all figures.

\subsection{Exact Finite-Shot Certification and Calibration Regret}

Certification at a declared target consumes an integer number of shots, and the sub-exponential prefactor of the multi-shot error probability is not negligible at practical targets. Figure~\ref{fig:workflow} summarizes how the three integration-time prescriptions used below --- $\tau_C$, $\tau_{\mathrm{finite}}(\epsilon)$ and $\tau_{\mathrm{BR}}$ --- follow from the score distributions and the per-shot overhead. Let $P_{\mathrm{e}}^{(N)}(\tau)$ denote the exact $N$-shot error probability of the optimal log-likelihood-ratio test built from $p_g$ and $p_e$ at integration time $\tau$. The required shot count and the exact certification time are
\begin{equation}
    N_{\min}(\tau;\epsilon) = \min\{N \in \mathbb{N} : P_{\mathrm{e}}^{(N)}(\tau) \le \epsilon\},
    \label{eq:nmin}
\end{equation}
\begin{equation}
    T_{\mathrm{exact}}(\tau;\epsilon) = N_{\min}(\tau;\epsilon)\,(\tau+\tau_{\text{oh}}),
    \label{eq:t_exact}
\end{equation}
so that $T_{\mathrm{exact}}$ is piecewise linear in $\tau$ between integer-shot thresholds and changes discontinuously whenever $N_{\min}$ changes. Writing $\tau_{\mathrm{finite}}(\epsilon) = \arg\min_{\tau} T_{\mathrm{exact}}(\tau;\epsilon)$, we quantify the relative excess certification time incurred by committing to a fixed, $\epsilon$-independent integration time $\tau$. We refer to this dimensionless excess cost as calibration regret:
\begin{equation}
    \mathcal{L}(\tau;\epsilon)
    =
    \frac{T_{\mathrm{exact}}(\tau;\epsilon)}
         {T_{\mathrm{exact}}(\tau_{\mathrm{finite}}(\epsilon);\epsilon)}
    - 1 .
    \label{eq:regret}
\end{equation}
Thus, $\mathcal{L}=0.10$, for example, means that the fixed calibration requires $10\%$ more wall-clock time than the exact target-specific optimum. Because $N_{\min}$ is integer-valued, no fixed rule can maintain uniformly small relative excess certification time: threshold mismatches between the fixed rule and the target-specific optimum produce calibration-regret spikes of order $O(1/N)$.

The sub-exponential structure of $P_{\mathrm{e}}^{(N)}$ is captured by the Bahadur--Ranga Rao sharp large-deviation refinement (hereafter Bahadur--Rao), applied here to sums of independent log-likelihood ratios \cite{bahadur1960deviations,efron1968large}:
\begin{equation}
    P_{\mathrm{e}}^{(N)}(\tau)
    \simeq
    \frac{A(\tau)}{\sqrt{N}}\,e^{-N C(\tau)},
    \label{eq:bahadur_rao}
\end{equation}
in which the prefactor $A(\tau)$ is fixed by the exponentially tilted log-likelihood-ratio distribution at the optimal Chernoff parameter $s^*$. Replacing $\log(1/\epsilon)/C(\tau)$ in Eq.~\eqref{eq:t_cert} by the continuous shot count that solves Eq.~\eqref{eq:bahadur_rao} at $P_{\mathrm{e}}^{(N)}=\epsilon$ defines a prefactor-corrected surrogate objective whose minimizer we denote $\tau_{\mathrm{BR}}(\epsilon)$. Because $\tau_{\mathrm{BR}}$ depends on the declared design target, it is not target-free. We therefore specify a practical target interval and a set of candidate values of $\epsilon_{\mathrm{design}}$ within it. The Bahadur--Rao-based surrogate objective is evaluated at each candidate. Section~III.E shows that the minimax and minimum-median criteria both select the shallowest candidate, $\epsilon_{\mathrm{design}}=10^{-4}$, at every surveyed parameter point.

Exact $N$-shot error probabilities are obtained by numerical convolution of the log-likelihood-ratio density, with exponential tilting for stability in the far tail, and the engine is validated against closed-form Gaussian results. The location of the continuous-$\tau$ finite-shot optima, the resolution of integer-shot transitions, and the associated numerical error bounds are described in the Supplemental Material.

\subsection{Asymptotic Scaling and Overhead-Induced Shift}

To clarify the structural origin of the systematic separation between $\tau_{\text{fid}}$ and $\tau_{C}$, we consider the long-coherence limit ($T_1 \rightarrow \infty$) where the matched-filter score distributions remain symmetric Gaussians. In this regime, the classical Chernoff information reduces to
\begin{equation}
C(\tau)=\frac{SNR^2(\tau)}{8}.
\end{equation}
The asymptotic objective \eqref{eq:t_cert} is therefore equivalent, up to a constant prefactor, to minimizing
\begin{equation}
\bar{T}_{\text{cert}}(\tau)\propto \frac{\tau+\tau_{\text{oh}}}{SNR^2(\tau)}.
\end{equation}
A stationary point satisfies the exact condition
\begin{equation}
\begin{aligned}
\frac{d}{d\tau}\!\left(\frac{\tau+\tau_{\text{oh}}}{\mathrm{SNR}^2(\tau)}\right) &= 0 \\
\Longleftrightarrow\quad
\mathrm{SNR}^2(\tau)
&= \bigl(\tau+\tau_{\text{oh}}\bigr)
\frac{d}{d\tau}\mathrm{SNR}^2(\tau).
\end{aligned}
\label{eq:tau_rate_condition}
\end{equation}

For dispersive readout with finite resonator linewidth, $\mathrm{SNR}^2(\tau)$ grows convexly in $\tau$: its derivative, proportional to $(1-e^{-\kappa\tau/2})^2$, increases monotonically from zero toward the steady-state information rate. Convex information accumulation alone, however, cannot produce an interior rate optimum. For any differentiable convex function $f(\tau)=\mathrm{SNR}^2(\tau)$ with $f(0)=0$, convexity gives $f(\tau)\leq \tau f'(\tau)$, with strict inequality when $f$ is strictly convex on $[0,\tau]$. Consequently, for any $\tau_{\text{oh}}>0$, one has $f(\tau)<(\tau+\tau_{\text{oh}})f'(\tau)$, so the stationarity condition \eqref{eq:tau_rate_condition} admits no solution. For $\tau_{\text{oh}}=0$, the strict convexity of the present cavity-ring-up function likewise excludes an interior solution. Thus, in the strict Gaussian limit, $\bar{T}_{\text{cert}}(\tau)$ decreases monotonically. The nominal optimum then moves to arbitrarily long integration windows. At sufficiently long times, the continuous large-shot proxy predicts fewer than one shot and loses its operational meaning. An interior optimum $\tau_C$ therefore requires the per-shot information to accumulate sublinearly at long times. In the physical model, $T_1$-induced trajectory smearing provides this saturation and caps $C(\tau)$ far below the Gaussian benchmark [Fig.~\ref{fig:smear_asymmetry}(c)]. Decoherence therefore creates the interior rate optimum, whereas the response of that optimum to hardware overhead is structural.

The overhead-induced shift follows from the corresponding stationarity condition for the physical Chernoff information, $C(\tau_C)=(\tau_C+\tau_{\text{oh}})C'(\tau_C)$. In the saturating, locally concave regime containing the optimum, $C'(\tau_C)>0$ and $C''(\tau_C)<0$. Implicit differentiation gives
\begin{equation}
\frac{d\tau_C}{d\tau_{\text{oh}}}
=
-\frac{C'(\tau_C)}
{(\tau_C+\tau_{\text{oh}})C''(\tau_C)}
>0,
\end{equation}
so increasing $\tau_{\text{oh}}$ pushes $\tau_C$ toward larger integration times. This is the expected amortization effect: when fixed temporal costs are nonzero, it becomes advantageous to extract more information per shot. At the standard-linewidth baseline, the minimum remains unique and $\tau_C$ increases monotonically over the range $\tau_{\mathrm{oh}}=0$--$30~\mu\mathrm{s}$ examined here (Supplemental Material, Sec.~\ref{sec:fast_reset}). Over the main-text range $\tau_{\mathrm{oh}}=5$--$30~\mu\mathrm{s}$, the same overhead trend is robust to variation of $\kappa T_1$ over three orders of magnitude.

\section{Results}

We use a dispersive shift $\chi=7.54~\mu\text{s}^{-1}$ ($\chi/2\pi=1.2~\text{MHz}$), a cavity linewidth $\kappa=31.42~\mu\text{s}^{-1}$ ($\kappa/2\pi=5.0~\text{MHz}$), and an average steady-state photon number $\bar n=80$. These values give $\chi/\kappa=0.24$ and a cavity ring-up timescale $2/\kappa=0.064~\mu\text{s}$. Unless otherwise stated, we assume a detection efficiency $\eta = 0.45$ and a qubit relaxation time $T_1 = 30~\mu\text{s}$. The ratio $\chi/\kappa \approx 0.2$--$0.3$ is comparable to that employed in high-fidelity readout experiments \cite{walter2017rapid, jeffrey2014fast}, and the absolute rates adopted here correspond to a standard-linewidth readout cavity. A second parameter set, $\chi/2\pi \approx 0.19$~MHz and $\kappa/2\pi \approx 0.80$~MHz at the same $\chi/\kappa$, is reported as a separate sub-MHz-linewidth regime study in the Supplemental Material. Because $T_1$ and $\tau_{\mathrm{oh}}$ are held fixed in physical units, the two regimes have different values of $\kappa T_1$ and $\kappa\tau_{\mathrm{oh}}$ and therefore are not related by a simple rescaling. More generally, $\mathrm{SNR}^2(\tau) = \eta\bar{n}\,(4\chi/\kappa)^2\,g(\kappa\tau)$, with the dimensionless ring-up function $g(y)=\int_0^{y}(1-e^{-u/2})^2\,du$, and the structure of the results is controlled by the dimensionless combinations $\chi/\kappa$, $\kappa T_1$, and $\kappa\tau_{\mathrm{oh}}$. Thus, even when $\chi/\kappa$ is unchanged, changing the absolute linewidth while holding $T_1$ and $\tau_{\mathrm{oh}}$ fixed changes the latter two controlling parameters. Average photon numbers $\bar{n}\sim50$--$100$ and measurement efficiencies $\eta\approx0.4$--$0.6$ are consistent with modern high-efficiency readout implementations \cite{reed2010high, rosenthal2021efficient}, while relaxation times $T_1\sim20$--$50~\mu\text{s}$ are typical of contemporary transmon platforms \cite{barends2013coherent}.

The assumed per-shot overhead $\tau_{\text{oh}} = 15~\mu\text{s}$ is defined as a lumped cycle cost incurred once for every independent sample, including state preparation or reset, resonator depletion, classical control processing, and re-arming of the measurement sequence. With this convention, $N$ repeated acquisition cycles require the wall-clock time $N(\tau+\tau_{\text{oh}})$ used in Eqs.~\eqref{eq:t_cert} and \eqref{eq:t_exact}. Experimental studies of active resonator reset and measurement-feedback cycles demonstrate that cavity depletion and control-flow latency can occupy a substantial fraction of the measurement cycle, particularly in architectures prioritizing Purcell protection and coherence \cite{bultink2016active, riste2012feedback}. While optimized fast-reset protocols can reduce this overhead, realistic superconducting control stacks commonly retain microsecond-scale latencies, motivating the inclusion of an explicit $\tau_{\text{oh}}$ term in throughput analysis.

\subsection{Information Accumulation and Trajectory Smearing}

\begin{figure*}[t]
    \centering
    \includegraphics[width=\textwidth]{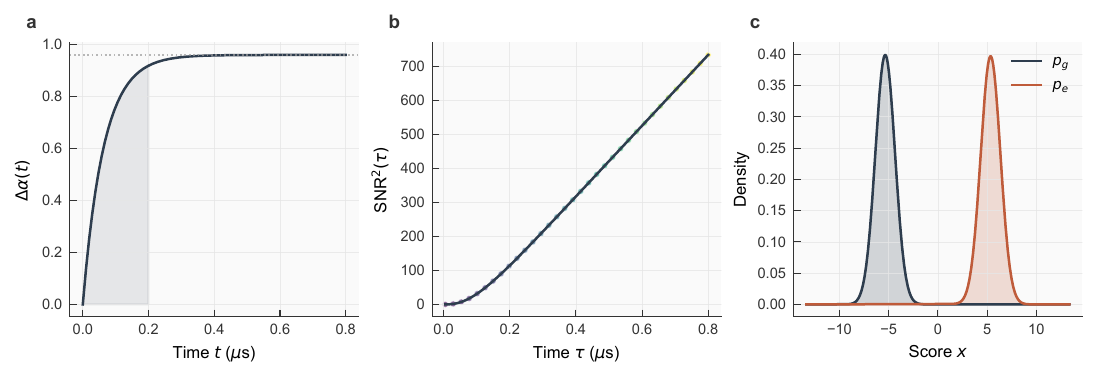}
\caption{\textbf{Dynamics of information accumulation in dispersive readout.} (a) Temporal evolution of the normalized branch-response function $\Delta\alpha(t)$ following the onset of the drive pulse; the two measured-quadrature branches sit at $\pm\Delta\alpha(t)$. The branch response approaches the steady-state limit $4\chi/\kappa$ as the resonator rings up. (b) Integrated signal-to-noise ratio squared ($SNR^2$) as a function of integration time $\tau$. Within the response model of Eq.~\eqref{eq:branch} the initially superlinear growth satisfies $SNR^2(\tau)\propto\tau^3$ at short times before approaching a linear-growth regime; the exact driven-cavity response gives $SNR^2(\tau)\propto\tau^5$ there, because the branch separation requires both field build-up and phase accumulation (Appendix~\ref{app:response}). (c) Score distributions $p_g(x)$ and $p_e(x)$ at $\tau=0.20~\mu$s. The excited-state distribution contains a small relaxation-induced component toward the ground-state mean; its information-theoretic effect is quantified in Fig.~\ref{fig:smear_asymmetry}.}
    \label{fig:physical_model}
\end{figure*}

\begin{figure*}[t]
    \centering
    \includegraphics[width=\textwidth]{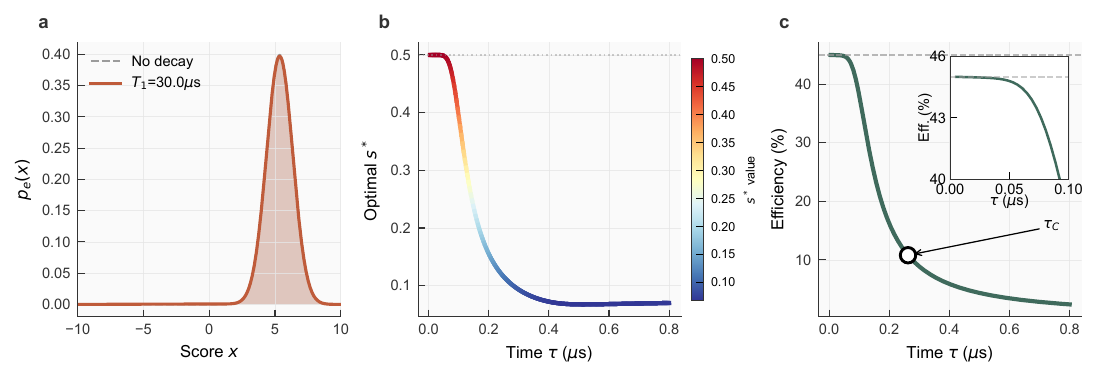}
 \caption{\textbf{Characterizing T$_1$-induced asymmetry and extraction efficiency.} (a) Comparison of the excited-state score distribution $p_e(x)$ in the ideal limit (no decay) and the realistic regime ($T_1 = 30~\mu$s). The relaxation trajectories broaden the excited-state score distribution toward the ground-state manifold; the resulting asymmetry is quantified by the optimal Chernoff parameter in panel (b). (b) The optimal Chernoff parameter $s^*$ as a function of integration time. At short $\tau$, the distributions remain nearly symmetric ($s^* \approx 0.5$), but as decay events accumulate, $s^*$ shifts toward zero, reflecting the increasing asymmetry of the two score distributions and their Chernoff overlap. (c) Information extraction efficiency $\eta_{\text{info}}$ relative to the unit-efficiency Gaussian Chernoff benchmark. The efficiency reaches its theoretical maximum of 45\% (limited by detection efficiency $\eta$) at short integration times and diminishes monotonically as $\tau$ increases, reflecting the progressive information loss driven by stochastic T$_1$ relaxation during the measurement window.}
    \label{fig:smear_asymmetry}
\end{figure*}

The dynamics of information accumulation are determined by the cavity response and the subsequent smearing of score distributions due to qubit decay. Figure~\ref{fig:physical_model}(a) shows the growth of the normalized branch-response function $\Delta\alpha(t)$, which approaches its steady-state value on a timescale governed by $\kappa^{-1}$. The integrated information, quantified by the squared signal-to-noise ratio $SNR^2(\tau)$, scales non-linearly during the ring-up phase before entering a linear growth regime [Fig.~\ref{fig:physical_model}(b)]. At $\tau=0.20~\mu$s, the resulting score distributions $p_g(x)$ and $p_e(x)$ are clearly separated. Relaxation contributes a small component toward the ground-state mean, whose information-theoretic effect is quantified more sensitively through the Chernoff asymmetry and extraction efficiency in Fig.~\ref{fig:smear_asymmetry}.

One consequence of this $T_1$-induced smearing is the introduction of asymmetry between the hypotheses. Figure~\ref{fig:smear_asymmetry}(a) compares the excited-state distribution with and without relaxation. The decay trajectories broaden it toward the ground-state manifold, which drives the optimal Chernoff parameter $s^*$ away from the symmetric Gaussian value $1/2$ as $\tau$ increases [Fig.~\ref{fig:smear_asymmetry}(b)]. The shift reflects the unequal weighting of the two hypotheses in the exponentially tilted distribution.

We quantify the quality of the measurement scheme via the information extraction efficiency $\eta_{\mathrm{info}}$, defined as the ratio of the classical Chernoff information $C(\tau)$ to the unit-efficiency Gaussian benchmark $C_{\mathrm{ideal}}(\tau)$. As shown in Fig.~\ref{fig:smear_asymmetry}(c), the efficiency reaches its maximum value of $45\%$ at short integration times, reflecting the finite detection efficiency $\eta$. As $\tau$ increases, the accumulated probability of $T_1$ relaxation events causes the classical information to deviate from the ideal Gaussian scaling, with the extraction efficiency falling to $\eta_{\mathrm{info}}(\tau_{C}) \approx 10.8\%$ at the rate-optimal integration time.

\subsection{Separation of Optimal Integration Times}

\begin{figure}[h]
    \centering
    \includegraphics[width=\columnwidth]{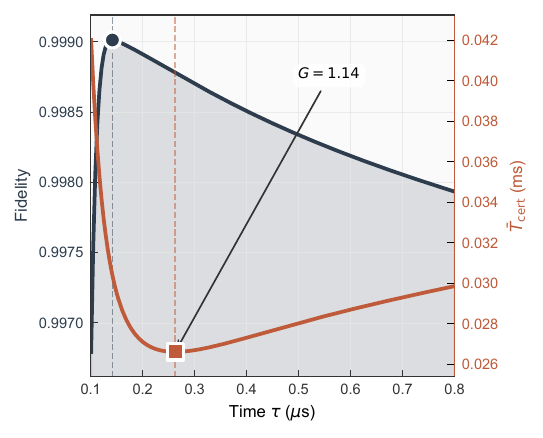}
   \caption{\textbf{Systematic separation between the fidelity and asymptotic-rate optima.} The Bayes fidelity (primary axis) and the asymptotic certification-time proxy $\bar{T}_{\text{cert}}(\tau)$ of Eq.~\eqref{eq:t_cert} (secondary axis) are plotted against integration time $\tau$, with the proxy evaluated at $\epsilon = 10^{-4}$. Vertical dashed lines (with corresponding markers) indicate the fidelity-optimal point ($\tau_{\text{fid}}$) and the rate-optimal point ($\tau_{C}$). Because the target error enters $\bar{T}_{\text{cert}}$ only through the prefactor $\log(1/\epsilon)$, the location of $\tau_{C}$ is exactly independent of $\epsilon$. The minimum of the proxy occurs at a noticeably longer integration window, demonstrating that throughput optimization involves a fundamental trade-off between single-shot confidence and per-shot hardware overhead. The exact finite-shot optimum is target-dependent and lies below $\tau_{C}$; it is not shown here and is quantified in Secs.~III.B and III.E.}
    \label{fig:divergence}
\end{figure}

The main observation is the systematic separation between the integration time that maximizes single-shot fidelity, $\tau_{\text{fid}}$, and the time that maximizes the asymptotic Chernoff rate, $\tau_{C}$. Figure~\ref{fig:divergence} illustrates this effect at the baseline parameters. While the Bayes fidelity peaks at $\tau_{\text{fid}} \approx 0.141~\mu\text{s}$, the Chernoff rate $\mathcal{R}_C(\tau)$ of Eq.~\eqref{eq:tau_C} attains its maximum at a significantly longer integration time, $\tau_{C} \approx 0.263~\mu\text{s}$, approximately $86\%$ longer.

This shift arises because the asymptotic objective $\bar{T}_{\text{cert}}$ accounts for both the information rate $C(\tau)$ and the fixed hardware overhead. At the rate-optimal point, integrating longer extracts more information per shot and maximizes the asymptotic Chernoff information accumulated per unit wall-clock time, rather than maximizing single-shot fidelity. At the baseline parameters, the asymptotic rate-gain factor is $G\approx1.14$. Across the ten-point practical survey in this standard-linewidth regime, $G$ ranges from $1.128$ to $1.157$, corresponding to percentage improvements of $12.8\%$--$15.7\%$; $G$ is exactly independent of the target error $\epsilon$.

The exact finite-shot objective behaves differently. Because $N_{\min}(\tau;\epsilon)$ of Eq.~\eqref{eq:nmin} is integer-valued, $T_{\mathrm{exact}}(\tau;\epsilon)$ is piecewise linear between shot-count thresholds and changes discontinuously when $N_{\min}$ changes; its minimizer therefore depends on the target. At the baseline parameters the refined target-specific optima are $\tau_{\mathrm{finite}} = 0.102~\mu$s ($N_{\min} = 2$) at $\epsilon = 10^{-4}$, $0.115~\mu$s ($N_{\min} = 4$) at $10^{-8}$, and $0.163~\mu$s ($N_{\min} = 5$) at $10^{-12}$: all lie \emph{below} $\tau_{C} = 0.263~\mu$s. Across all twenty surveyed parameter points, the refined exact optimum satisfies $\tau_{\mathrm{finite}}(\epsilon)\le\tau_C$. The endpoint trend $\tau_{\mathrm{finite}}(10^{-12}) \ge\tau_{\mathrm{finite}}(10^{-4})$ also holds at every point. This behavior is consistent with the Bahadur--Rao correction of Eq.~\eqref{eq:bahadur_rao}, which returns the optimum to $\tau_C$ only as $\epsilon\to0$. The progression with target stringency is sawtooth-like rather than strictly monotonic; the refined optimum shifts, for example, from approximately $0.164~\mu$s at $\epsilon = 10^{-10}$ to $0.163~\mu$s at $10^{-12}$.

\subsection{Scaling of the Asymptotic Rate Gain}

\begin{figure*}[t]
    \centering
    \includegraphics[width=\textwidth]{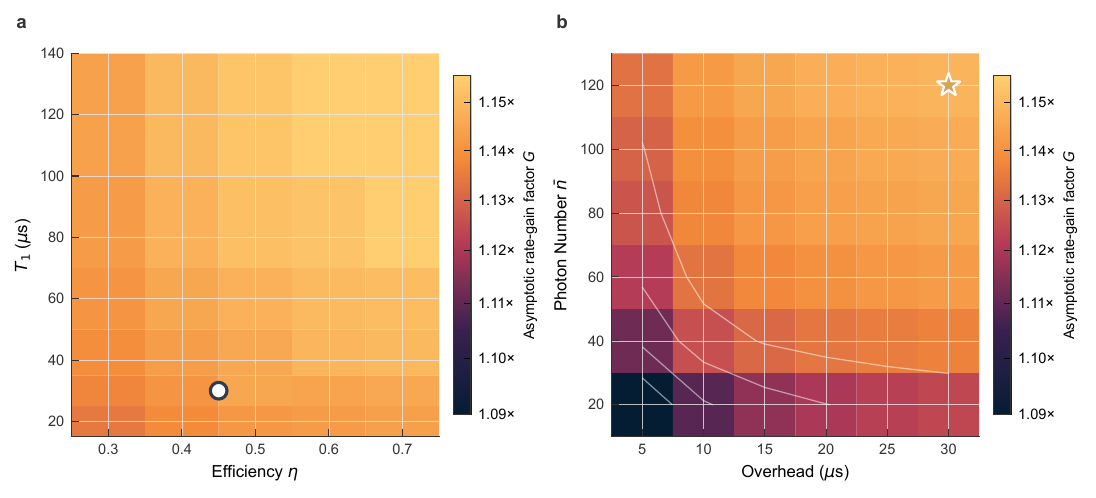}
\caption{\textbf{Scaling of the asymptotic rate gain across parameter space.} (a) Asymptotic rate gain $G$ as a function of qubit relaxation time $T_1$ and detection efficiency $\eta$. Higher efficiencies allow the measurement to reach the certification threshold more rapidly, making the fixed hardware overhead the dominant bottleneck. The white circle marks the baseline parameter location ($T_1=30~\mu$s, $\eta=0.45$) used throughout this work; the value $\eta=0.45$ lies between the sampled efficiency nodes. (b) Asymptotic rate gain as a function of steady-state photon number $\bar{n}$ and per-shot hardware overhead, evaluated at $T_1=20~\mu$s and $\eta=0.5$. The curved white lines in this panel are contours of constant $G$, located by interpolation between neighboring points of the numerically evaluated parameter grid; the contour levels lie below the range of $G$ spanned by panel (a), so no contour appears there. Straight white lines in both panels are axis gridlines. The gold star marks the largest gain within the panel-(b) sweep. The heatmap sweeps are distinct from the ten-point practical survey quoted in the text and are not constrained to its gain range. Both panels use a common color scale spanning the combined calculated range of the two sweeps. The quoted values are asymptotic rate gains rather than realized finite-target time reductions, and no universal ceiling is implied by the plotted range.}
    \label{fig:heatmaps}
\end{figure*}

The magnitude of the asymptotic rate gain depends on the interplay between qubit coherence and hardware constraints. Figure~\ref{fig:heatmaps}(a) maps the rate gain $G$ across a range of $T_1$ times and detection efficiencies $\eta$. At fixed measurement efficiency, longer $T_1$ produces a modest increase in the gain, although the dependence on $T_1$ is weaker than the dependence on $\eta$ or the per-shot overhead. Longer coherence reduces decay-induced asymmetry, which allows more information to be accumulated per shot.

As defined above, the per-shot overhead $\tau_{\text{oh}}$ is a lumped cycle cost incurred once for every independent sample. We take $\tau_{\mathrm{oh}} = 15~\mu$s as a representative mid-range value for the baseline results. Systems combining fast active reset with low-latency resonator depletion, control processing, and sequence re-arming can have total per-shot overheads below this range. The behavior of the optimum for $\tau_{\mathrm{oh}}$ down to zero is reported in Sec.~\ref{sec:fast_reset} of the Supplemental Material. We examine the range $\tau_{\mathrm{oh}}=5$--$30~\mu$s in two complementary settings: Fig.~\ref{fig:heatmaps}(b) maps its influence in the practical finite-coherence regime at $T_1=20~\mu$s, whereas Fig.~\ref{fig:validation}(b) repeats the overhead sweep in the long-coherence regime to isolate the structural overhead-induced shift. This range represents cycles built on active reset or active state preparation, for which demonstrated reset times reach the few-hundred-nanosecond scale \cite{Magnard2018fast,krinner2022realizing} and the remaining cycle cost is dominated by resonator depletion, classical processing, and re-arming. It is not intended to represent passive thermalization, which for $T_1\sim20$--$50~\mu$s requires waiting several multiples of $T_1$ and yields a substantially larger $\tau_{\mathrm{oh}}$. The objective of Eq.~\eqref{eq:t_cert} is unchanged in that case: one inserts the measured cycle overhead. The independent-shot assumption requires only that the chosen reset or preparation protocol restore the intended initial state with sufficient reliability.

The photon-number sweep in Fig.~\ref{fig:heatmaps}(b) is interpreted within the validity of the reduced dispersive and linear-response model. In a specific device, the admissible upper range of $\bar n$ must additionally satisfy hardware-dependent constraints associated with the critical photon number, transmon ionization, and other nonlinear readout effects, which are not included here.
\subsection{Consistency in the Gaussian Limit}

\begin{figure*}[t]
    \centering
    \includegraphics[width=\textwidth]{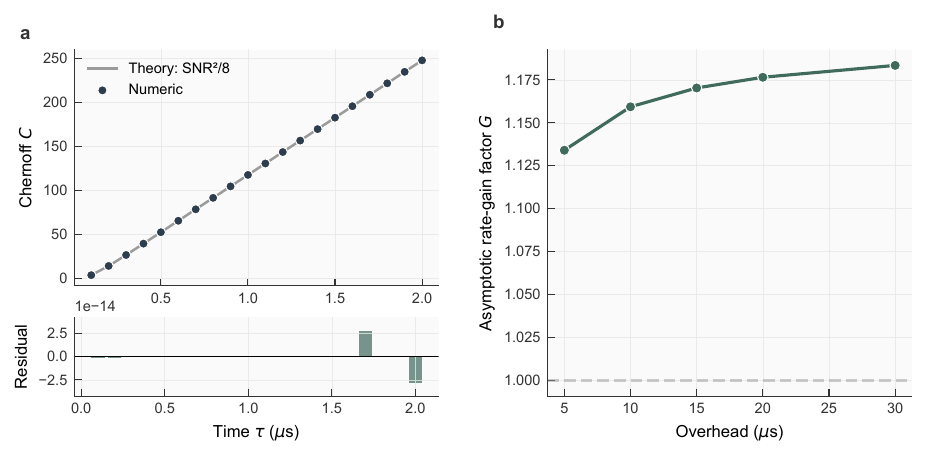}
    \caption{\textbf{Framework validation and the Gaussian limit.} (a) Calculated Chernoff information $C$ (points) compared to the analytical $SNR^2/8$ limit (line) in the absence of qubit decay ($T_1 \to \infty$). The lower panel shows residuals ($C_{\mathrm{num}}-C_{\mathrm{theory}}$) on a $10^{-14}$ scale, at the level of double-precision round-off for $C\sim10^{2}$, confirming the score-grid quadrature and Chernoff-parameter optimization in the Gaussian limit. (b) Asymptotic rate gain versus per-shot overhead in the long-coherence regime ($\kappa T_1 = 5\times10^{6}$). The overhead axis spans the main-text survey range; the extension to $\tau_{\mathrm{oh}}=0$ at the standard-linewidth baseline is reported in Sec.~\ref{sec:fast_reset}. Over the range shown, the gain increases monotonically with overhead, consistent with the amortization argument of Sec.~II.E: the rate optimum $\tau_{C}$ shifts to longer integration times while the fidelity optimum $\tau_{\mathrm{fid}}$ is overhead-independent. This long-coherence diagnostic is separate from the ten-point practical survey quoted in the main text.}
    \label{fig:validation}
\end{figure*}

We validate the score-distribution integration and optimization over the Chernoff parameter in the limit of infinite qubit coherence, obtained by switching off the relaxation channel exactly, in which case the score distributions are symmetric Gaussians. Figure~\ref{fig:validation}(a) shows that the numerical Chernoff information agrees with the analytical Gaussian result $SNR^2/8$ to machine precision, confirming the score-grid quadrature and the optimization over the Chernoff parameter. Because the relaxation channel vanishes in this limit, this panel does not test the decay-time quadrature.

Figure~\ref{fig:validation}(b) probes the long-coherence regime, evaluated at $\kappa T_1 = 5\times10^{6}$ so that relaxation is negligible within any single integration window. Even here, the residual decay channel governs both optima. The Bayes error is the sum of a Gaussian-overlap contribution, which falls steeply with $\tau$, and a decay-induced contribution growing as $\tau/T_1$; their crossover fixes $\tau_{\mathrm{fid}}=0.213~\mu$s, later than the $\tau_{\mathrm{fid}}\approx 0.141~\mu$s of Fig.~\ref{fig:divergence} and drifting only logarithmically with $T_1$. Likewise, it is the saturation of $C(\tau)$ by the residual decay tail that furnishes the interior minimum of $\bar{T}_{\text{cert}}$, consistent with the analysis of Sec.~II.E. Over the range $\tau_{\mathrm{oh}}=5$--$30~\mu$s shown here, the rate gain grows monotonically with overhead, and this growth is insensitive to the residual-decay scale: repeating the sweep with $\kappa T_1$ varied over three orders of magnitude changes individual gain values by less than $0.01$ per decade while preserving the trend. The gain is not monotonic over the wider range extending to $\tau_{\mathrm{oh}}=0$, as shown in Sec.~\ref{sec:fast_reset}. This separation reflects the difference between the single-shot Bayes-fidelity objective and the multi-shot Chernoff error-exponent objective. Increasing the hardware overhead $\tau_{\mathrm{oh}}$ further amplifies this effect over the physically relevant range, with the asymptotic rate gain growing from $1.134\times$ to $1.183\times$ as $\tau_{\mathrm{oh}}$ increases from $5$ to $30~\mu$s. This long-coherence diagnostic is distinct from the ten-point practical standard-linewidth survey, whose asymptotic gains range from $12.8\%$ to $15.7\%$; the larger value reached here therefore does not extend the quoted practical-survey range.

These calculations separate the two roles: relaxation-induced saturation creates the interior rate optimum, while the overhead term determines its location.

\subsection{Finite-Shot Calibration Regret and the Bahadur--Rao Preset}

We now evaluate fixed integration-time rules against the exact finite-shot optimum of Eq.~\eqref{eq:t_exact}, using the relative excess certification time (calibration regret) defined in Eq.~\eqref{eq:regret}, over a declared target interval with $\log_{10}(1/\epsilon)$ sampled uniformly on $[3,12]$. Interval widths are quoted in dex, where one dex denotes one unit of $\log_{10}(1/\epsilon)$. All quoted fractions of the interval are descriptive fractions of this logarithmic range and are not probabilities assigned to user targets. For summary statistics, we partition the range into the four fixed logarithmic bands $[3,5]$, $[5,7]$, $[7,9]$, and $[9,12]$; the worst-band median is the largest median regret among these four bands. The survey comprises twenty parameter points, ten in the standard-linewidth regime and ten in the sub-MHz-linewidth regime reported in the Supplemental Material.

At every surveyed point, a contiguous band of fixed integration times improves upon $\tau_C$ simultaneously in overall median regret, maximum regret, the fraction of the target interval exceeding $5\%$ regret, and worst-band median regret. This multi-metric dominant band lies strictly between $\tau_{\mathrm{fid}}$ and $\tau_C$. The band is not itself a calibration rule, and the performance of any particular preset is established from exact finite-shot regret tables rather than inferred from proximity to the band.

The rate-optimal time $\tau_C$ performs well on typical targets but poorly in localized threshold regions. Its median regret ranges from $0.35\%$ to $5.28\%$ across the twenty points, while its maximum regret ranges from $50.5\%$ to $104.2\%$. The fraction of the target interval on which $\tau_C$ exceeds $5\%$ regret ranges from $1.9\%$ to $6.8\%$ at nineteen of twenty points. At the twentieth, the low-overhead point of the sub-MHz survey ($\tau_{\mathrm{oh}}=5~\mu$s), it reaches $55.6\%$. The widest such window spans $1.38$ dex there, against $0.057$--$0.351$ dex elsewhere. Statements concerning narrow high-regret windows must therefore be conditioned on $\kappa\tau_{\mathrm{oh}}$.

Evaluating the Bahadur--Rao-based surrogate objective at the shallowest candidate design target, $\epsilon_{\mathrm{design}}=10^{-4}$, gives $\tau_{\mathrm{BR}}=0.221~\mu$s at the standard-linewidth baseline and $1.055~\mu$s at the sub-MHz baseline. Among the candidate design targets $\{10^{-4},10^{-8},10^{-12}\}$, both the minimax and minimum-median criteria select $\epsilon_{\mathrm{design}}=10^{-4}$ at all twenty points, thereby fixing the prescription. Relative to $\tau_C$, the preset $\tau_{\mathrm{BR}}$ is superior at all twenty points in overall median, worst-band median, and maximum regret. Its overall median regret is at most $4.42\%$, and its worst-band median is at most $6.8\%$. The comparison with $\tau_{\mathrm{fid}}$ instead reflects a difference in risk profile. The preset lowers the worst-band median at nineteen of twenty points, while the corresponding values for $\tau_{\mathrm{fid}}$ reach $24.9\%$. However, $\tau_{\mathrm{fid}}$ retains the lower overall median at all twenty points because shallow targets dominate the logarithmic interval. The empirically minimum-median fixed time inherits the same stringent-target exposure as $\tau_{\mathrm{fid}}$, with worst-band median regret reaching $24.9\%$.

\begin{figure*}[t]
    \centering
    \includegraphics[width=\textwidth]
    {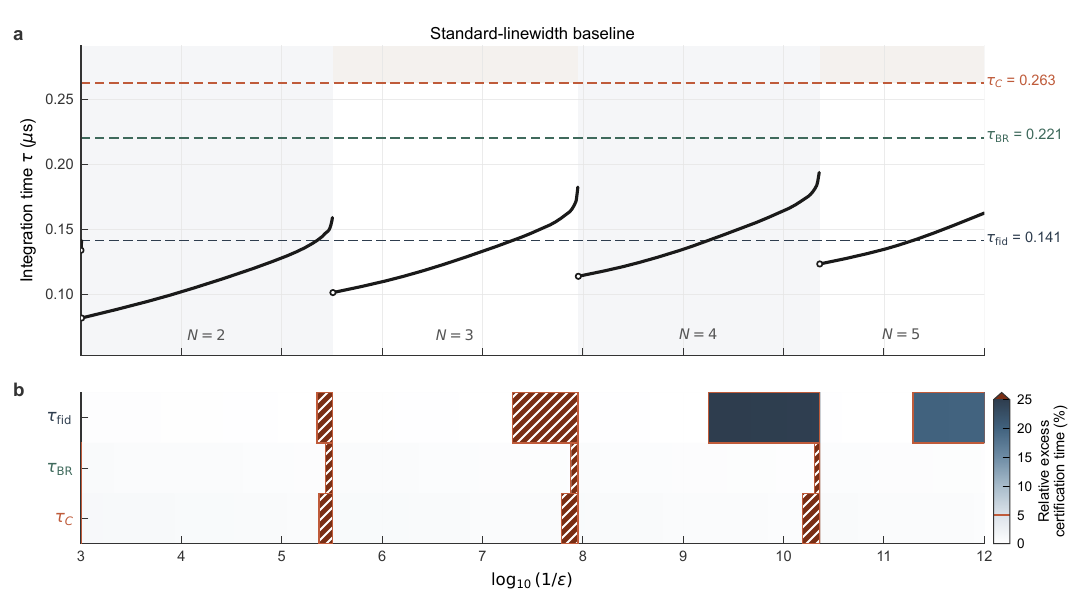}
    \caption{\textbf{Exact finite-shot calibration at the standard-linewidth baseline.} (a) Target-specific optimum $\tau_{\mathrm{finite}}(\epsilon)$ versus $\log_{10}(1/\epsilon)$. Alternating bands denote intervals of constant required shot count $N_{\min}$. Within each band the optimum varies continuously, while changes in $N_{\min}$ produce discontinuous resets, giving a sawtooth-like dependence that is not strictly monotonic. Horizontal lines mark the fidelity optimum $\tau_{\mathrm{fid}}$, the Bahadur--Rao preset $\tau_{\mathrm{BR}}$ evaluated at $\epsilon_{\mathrm{design}}=10^{-4}$, and the asymptotic rate optimum $\tau_C$. The exact optimum remains below $\tau_C$ throughout the displayed target range and converges to $\tau_C$ only in the asymptotic limit $\epsilon\to0$. (b) Relative excess certification time (calibration regret) of the three fixed rules, $100[T_{\mathrm{exact}}(\tau;\epsilon)/ T_{\mathrm{exact}}(\tau_{\mathrm{finite}}(\epsilon);\epsilon)-1]$. Outlined regions exceed $5\%$, while hatched regions exceed the displayed $25\%$ color range. Although all three rules have low median regret, each develops localized integer-threshold spikes; hence no fixed integration time is uniformly optimal over the full target interval.}
    \label{fig:finite_shot_calibration}
\end{figure*}

Figure~\ref{fig:finite_shot_calibration} illustrates the target dependence of the exact finite-shot optimum and the relative excess certification time of the three fixed calibration rules at the standard-linewidth baseline.

No fixed rule, including $\tau_{\mathrm{BR}}$, eliminates integer-shot threshold spikes. When a fixed rule requires $N+1$ shots while the exact optimum requires $N$, the corresponding regret is
\begin{equation}
\mathcal{L}
=
\frac{(N+1)(\tau_{\mathrm{rule}}+\tau_{\mathrm{oh}})}
{N[\tau_{\mathrm{finite}}(\epsilon)+\tau_{\mathrm{oh}}]}
-1 .
\end{equation}
It is therefore of order $1/N$ when the competing per-shot durations are close. Among the surveyed maximum-regret events attributable to $N=2\to3$ mismatches, the values span $50.0\%$--$56.2\%$, while the $N=3\to4$ cases give $33.3\%$ to the displayed precision. The upper end of the former range occurs at the low-overhead sub-MHz point, where the difference between $\tau_{\mathrm{rule}}$ and $\tau_{\mathrm{finite}}$ is appreciable relative to $\tau_{\mathrm{oh}}$. The observed values near $100$--$104\%$ occur when the exact optimum certifies in a single shot while the fixed rule requires two, either at a rule transition inside the declared interval or, at the standard-linewidth baseline, at its shallow endpoint. The rule $\tau_{\mathrm{fid}}$ is structurally exempt: it minimises $P_{\mathrm{e}}^{(1)}$ by construction, so $\lambda_1(\tau)=-\log_{10}P_{\mathrm{e}}^{(1)}(\tau)$ attains its global maximum there and no shorter integration time can certify in one shot at a target $\tau_{\mathrm{fid}}$ cannot. Its maximum regret is consequently $50.0\%$ at all twenty surveyed points. At each of the six standard-linewidth points where single-shot certification is feasible within the declared interval, both $\tau_{\mathrm{BR}}$ and $\tau_C$ exceed $100\%$ maximum regret; across those two rules the values span $100.6\%$--$104.2\%$. The empirically minimum-median fixed time reaches $100.2\%$ and $100.6\%$ at only two of those six points. At the other four it coincides with $\tau_{\mathrm{fid}}$ and therefore has the same $50.0\%$ maximum. Exact target-specific calibration, based on a precomputed finite-shot error map over candidate integration times, is therefore required to attain the minimum certification time at every specified target within the present model. When a specific target $\epsilon$ is known, $\tau_{\mathrm{finite}}(\epsilon)$ should be used; when only a practical target interval has been declared, the Bahadur--Rao preset evaluated at the shallowest candidate design target provides the validated fixed choice. The sub-MHz counterpart of Fig.~\ref{fig:finite_shot_calibration} and the full twenty-point fixed-rule comparison are provided in the Supplemental Material.

\begin{figure}[t]
\centering
\begin{tikzpicture}[
  bbpad,
  font=\footnotesize,
  box/.style={draw, rounded corners=2pt, align=center, inner sep=4pt,
              text width=58mm, minimum height=7mm},
  sel/.style={draw, rounded corners=2pt, align=center, inner sep=3pt,
              text width=20mm, minimum height=13mm, anchor=north},
  ar/.style={-{Stealth[length=2mm]}}
]
\node[box] (src) {Readout model (Sec.~II\,A--B) \emph{or} calibrated experimental I/Q data};
\node[box, below=5mm of src] (sc) {Score distributions $p_g(x;\tau)$, $p_e(x;\tau)$};
\node[box, below=5mm of sc] (ch) {Chernoff information $C(\tau)$; per-shot overhead $\tau_{\mathrm{oh}}$};
\node[box, below=5mm of ch] (sel) {Integration-time selection};
\node[sel, below left=7mm and 12mm of sel.south] (tc) {$\tau_C$\\[1pt] \scriptsize asymptotic rate optimum, Eq.~\eqref{eq:tau_C}};
\node[sel, below=7mm of sel.south] (tf) {$\tau_{\mathrm{finite}}(\epsilon)$\\[1pt] \scriptsize exact optimum at a known target};
\node[sel, below right=7mm and 12mm of sel.south] (tb) {$\tau_{\mathrm{BR}}$\\[1pt] \scriptsize preset for a declared target interval};
\draw[ar] (src) -- (sc);
\draw[ar] (sc) -- (ch);
\draw[ar] (ch) -- (sel);
\node[box, below=6mm of tf.south] (as) {Assessment against the exact finite-shot optimum: calibration regret $\mathcal{L}(\tau;\epsilon)$, Eq.~\eqref{eq:regret}};
\draw[ar] (sel.south) -- ++(0,-3mm) -| (tc.north);
\draw[ar] (sel.south) -- (tf.north);
\draw[ar] (sel.south) -- ++(0,-3mm) -| (tb.north);
\draw[ar] (tc.south) |- ++(0,-3mm) -| (as.north);
\draw[ar] (tb.south) |- ++(0,-3mm) -| (as.north);
\draw[ar] (tf.south) -- (as.north);
\end{tikzpicture}
\caption{\textbf{Calibration workflow.} The score distributions may come either from the readout model of Secs.~II\,A--B or directly from calibrated experimental I/Q data; the subsequent analysis is identical. Together with the per-shot overhead they determine three distinct integration-time prescriptions: the asymptotic rate optimum $\tau_C$ [Eq.~\eqref{eq:tau_C}], which is independent of the target error; the exact finite-shot optimum $\tau_{\mathrm{finite}}(\epsilon)$, used when a specific target $\epsilon$ is known; and the Bahadur--Rao preset $\tau_{\mathrm{BR}}$, a single fixed time selected when only a target interval has been declared. Every fixed rule is then assessed against the exact finite-shot optimum through the calibration regret of Eq.~\eqref{eq:regret}, evaluated from exact $N$-shot error probabilities over the declared target interval.}
\label{fig:workflow}
\end{figure}

\FloatBarrier
\section{Conclusion}

The optimization of dispersive readout has traditionally focused on single-shot distinguishability, while Chernoff information is already established as its generalization to repeated analog readout \cite{danjou2021generalized}. The present work addresses a different optimization problem: Chernoff information alone does not determine the wall-clock-optimal calibration when the integration time, fixed per-shot overhead, and integer number of shots required at a declared target error are included. By framing readout as a stochastic communication channel and optimizing the Chernoff error exponent directly, we identify a systematic separation between the fidelity-optimal integration time $\tau_{\text{fid}}$ and the rate-optimal time $\tau_{C}$.

The separation has two physical ingredients. Qubit relaxation caps the per-shot Chernoff information below its Gaussian benchmark. This saturation creates the interior minimum of the asymptotic objective; in the strict no-decay limit, the wall-clock objective instead decreases monotonically with integration time. Once the information channel is saturating and locally concave, any nonzero per-shot overhead $\tau_{\text{oh}}$ shifts the rate optimum toward longer integration windows. This is an amortization effect: when the fixed cost per shot is large, it is advantageous to extract more information from each shot.

For representative transmon parameters the rate-optimal window lies approximately $86\%$ beyond the fidelity-optimal time. The resulting asymptotic rate gain is exactly independent of the target error but depends on the parameter regime, ranging from $12.8\%$ to $15.7\%$ across the ten-point practical standard-linewidth survey and from $6.3\%$ to $11.6\%$ across the corresponding sub-MHz survey. At a specified finite target, the realized wall-clock gain is target-dependent and can differ substantially from the asymptotic rate gain because certification consumes an integer number of shots: the exact optimum lies at or below $\tau_{C}$ at every surveyed point and returns to it only asymptotically. A Bahadur--Rao preset, selected at the shallowest candidate design target of a declared interval, lowers both the overall median and the worst-band median calibration regret relative to $\tau_{C}$ at every surveyed point. Its cost is a higher overall median than $\tau_{\text{fid}}$. No fixed rule removes integer-shot regret spikes. Attaining the minimum certification time at every target therefore requires target-specific calibration from a precomputed finite-shot error map. The magnitude of the rate advantage depends strongly on the total per-shot overhead. At the standard-linewidth baseline, $\tau_C$ crosses $\tau_{\mathrm{fid}}$ near $\tau_{\mathrm{oh}}=0.10~\mu$s; below this crossing, the ordering reverses and $\tau_C<\tau_{\mathrm{fid}}$. For total overheads of $0.2$--$0.5~\mu$s, $\tau_C$ remains longer than $\tau_{\mathrm{fid}}$, but the asymptotic gain is only $0.8\%$--$4.0\%$, so the practical benefit of distinguishing the two calibrations is limited in that regime. The asymptotic-rate optimization, the fixed-preset strategy, and the target-specific calibration require only changes to the calibration procedure, not to the hardware. The gains quantified here apply when analog or soft readout information is retained across repeated shots; a binarized acquisition chain requires optimization of the corresponding binary channel (Sec.~II.C).

Benchmarking classical Chernoff extraction against the unit-efficiency Gaussian Chernoff benchmark, we propose a diagnostic for quantifying information loss that can be evaluated from measured score distributions. Extraction efficiency falls to $\eta_{\mathrm{info}}(\tau_{C}) \approx 10.8\%$ at the rate-optimal integration time, which separates the contribution of amplifier inefficiency from that of $T_1$-induced trajectory smearing.

As superconducting processors scale and repetition rates increase, readout and reset latency will become increasingly dominant contributors to runtime. This analysis reframes measurement calibration in terms of wall-clock certification time, aligning readout optimization with the operational objective of large-scale quantum processors.

\appendix
\section*{Supplemental Material}

The tables and figures below are assembled from the validated numerical outputs used in the main text and from the additional fast-reset scan reported in Sec.~\ref{sec:fast_reset}. No result is inferred from figure digitization or from unvalidated intermediate calculations. Symbols follow the main text: $\tau_{\mathrm{fid}}$ is the fidelity-optimal integration time, $\tau_{C}$ the rate-optimal time, $\tau_{\mathrm{BR}}$ the Bahadur--Rao preset, and $\tau_{\mathrm{finite}} (\epsilon)$ the exact target-specific optimum. Calibration regret $\mathcal{L}(\tau;\epsilon)$ is defined in Eq.~\eqref{eq:regret} of the main text.

\section{Sub-MHz-linewidth regime study}

This regime uses $\chi/2\pi \approx 0.19$~MHz and $\kappa/2\pi \approx 0.80$~MHz. It has the same ratio $\chi/\kappa = 0.24$ as the standard-linewidth baseline of the main text but, because $T_1$ and $\tau_{\mathrm{oh}}$ are held fixed in physical units, different values of the controlling dimensionless combinations $\kappa T_1$ and $\kappa\tau_{\mathrm{oh}}$. The two regimes are therefore distinct physical settings rather than a rescaling of one another.

At the sub-MHz baseline the optima are $\tau_{\mathrm{fid}} = 0.784~\mu$s and $\tau_{C} = 1.199~\mu$s, an asymptotic rate-gain factor of $G = 1.101$. Across the ten sub-MHz survey points the gain spans $6.3\%$--$11.6\%$, compared with $12.8\%$--$15.7\%$ in the standard-linewidth regime (Table~\ref{tab:points}). Evaluating the Bahadur--Rao-based surrogate objective at $\epsilon_{\mathrm{design}}=10^{-4}$ gives $\tau_{\mathrm{BR}}=1.055~\mu$s.

\begin{figure*}[t]
  \centering
  \includegraphics[width=\textwidth]{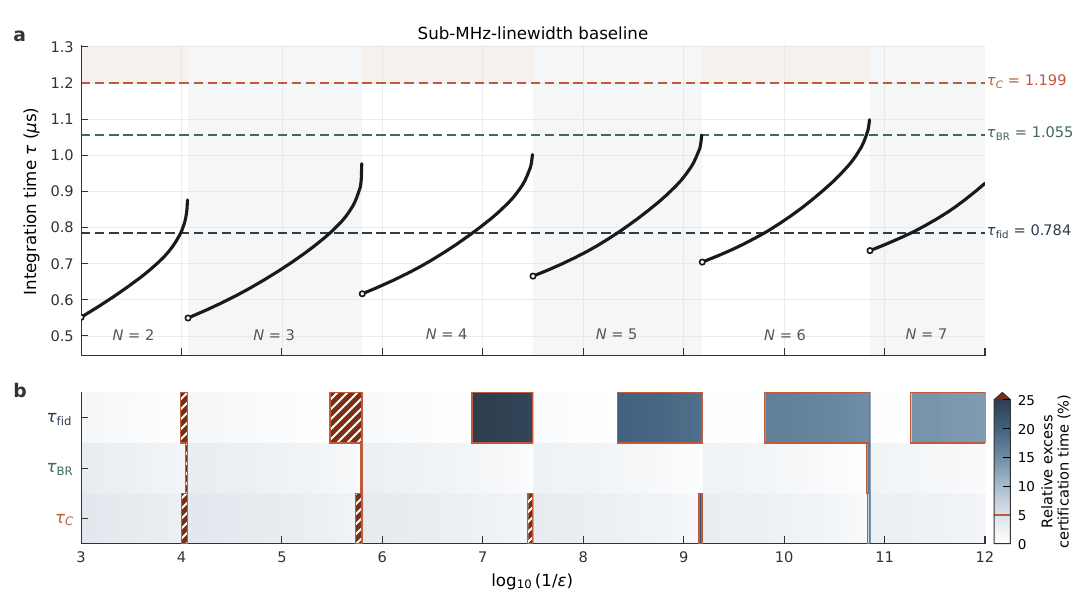}
    \caption{\textbf{Exact finite-shot calibration at the sub-MHz-linewidth baseline.} This figure is the sub-MHz counterpart of Figure~\ref{fig:finite_shot_calibration}  in the main text.  (a)~Target-specific optimum $\tau_{\mathrm{finite}}(\epsilon)$; alternating bands denote intervals of constant required shot count $N_{\min}$, and horizontal lines mark $\tau_{\mathrm{fid}}$, $\tau_{\mathrm{BR}}$ and $\tau_{C}$. (b)~Relative excess certification time of the three fixed rules; outlined regions exceed $5\%$ and hatched regions exceed the displayed $25\%$ colour range.}
  \label{fig:submhz}
\end{figure*}

Figure~\ref{fig:submhz} is the sub-MHz counterpart of Fig.~6 of the main text. The same structure holds: the exact optimum $\tau_{\mathrm{finite}}(\epsilon)$ is sawtooth-like in $\log_{10}(1/\epsilon)$, remains below $\tau_{C}$ throughout the displayed range, and is not strictly monotonic; the fixed rules are typically low-regret but develop localized integer-threshold spikes.

\section{Definitions of the finite-shot summary metrics}
\label{app:metrics}

All statistics are evaluated over the declared target interval with $\log_{10}(1/\epsilon)$ sampled uniformly on $[3,12]$. Quoted fractions are descriptive fractions of that logarithmic range and are not probabilities assigned to user targets. Interval widths are quoted in dex, one dex being one unit of $\log_{10}(1/\epsilon)$. For a fixed rule $\tau$:

\begin{itemize}
\item \emph{Overall median} --- the median of $\mathcal{L}(\tau;\epsilon)$ over the interval.
\item \emph{Worst-band median} --- the interval is partitioned into the four fixed bands $[3,5]$, $[5,7]$, $[7,9]$ and $[9,12]$; the worst-band median is the largest of the four within-band medians. It measures exposure to a contiguous sub-range of targets, which the overall median hides because shallow targets dominate the logarithmic interval.
\item \emph{Maximum regret} --- the supremum of $\mathcal{L}$ over the declared interval, taken as the larger of the maximum on the sampled target grid and the exact right-hand limits at the integer-shot breakpoints $\lambda_N=-\log_{10}P_{\mathrm{e}}^{(N)}(\tau)$ of the rule, so that threshold spikes are not missed by sampling. The supremum may also be attained at the shallow endpoint $\log_{10}(1/\epsilon)=3$ when a single shot suffices for the exact optimum but not for the fixed rule; this occurs for $\tau_{\mathrm{BR}}$ and $\tau_C$ at the standard-linewidth baseline. Because $\tau_{\mathrm{fid}}$ minimizes the single-shot error, $\lambda_1(\tau)$ is maximal there. Consequently, no alternative integration time can remain feasible with one shot immediately beyond the one-shot threshold of the $\tau_{\mathrm{fid}}$ rule.
\item \emph{Fraction above $5\%$} --- the fraction of the interval on which $\mathcal{L} > 0.05$.
\item \emph{Widest window} --- the width in dex of the widest contiguous interval on which $\mathcal{L} > 0.05$, with edges refined by bisection.
\end{itemize}

The \emph{empirically minimum-median} fixed time (``best fixed'' in the tables) is the candidate integration time minimising the overall median at that parameter point. Its candidate set is the coarse selection grid of the asymptotic sweep, augmented with the exact values $\tau_{\mathrm{fid}}$ and $\tau_{C}$, so that both named rules are always available as candidates. Selection is performed on that coarse grid, whereas the regret statistics reported in Tables~\ref{tab:regret_median} and \ref{tab:regret_max} are recomputed for the selected time on the fine reporting grid, which resolves the integer-shot breakpoints of each rule. The two grids are not required to agree, and at several points --- including the five visible in Table~\ref{tab:regret_median} (standard-linewidth P04, P05 and P08; sub-MHz P02 and P03) --- the fine-grid overall median of the selected time slightly exceeds that of $\tau_{\mathrm{fid}}$, by at most $0.12$ percentage points. The quantity is therefore a grid-restricted empirical minimiser, not a certified optimum over integration time, and it is reported only to show that optimising the median alone inherits the tail exposure of $\tau_{\mathrm{fid}}$; none of the conclusions depend on it. The selected integration time at each parameter point is reported in Table~\ref{tab:bestfixed}; at the standard-linewidth points at which single-shot certification is feasible within the declared interval it coincides with $\tau_{\mathrm{fid}}$ at four of six points, which is why the two rules share the same maximum regret there.

\section{Twenty-point parameter sweep and fixed-rule comparison}

\begin{figure}[t]
  \centering
  \includegraphics[width=\columnwidth]{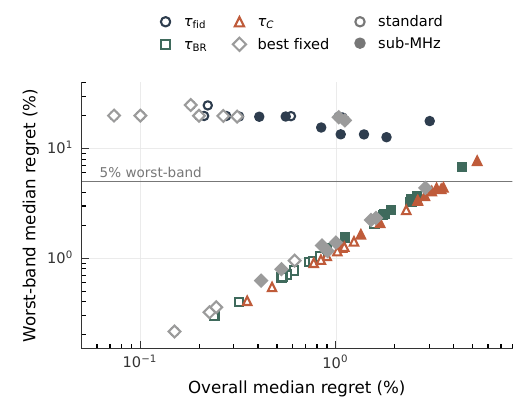}
  \caption{\textbf{Risk profile of the four fixed calibration rules across the twenty-point survey.} Overall median regret against worst-band median regret, one marker per parameter point and rule; open markers are standard-linewidth points, filled markers sub-MHz. Both axes are logarithmic. The fidelity-optimal time and the empirically minimum-median fixed time achieve the lowest overall median but scatter to worst-band medians as large as $24.9\%$, exceeding $5\%$ at $19$ and $8$ of the twenty points respectively. The Bahadur--Rao preset and the rate-optimal time trade a modestly higher overall median for a worst-band median bounded by $6.8\%$ and $7.7\%$, exceeding $5\%$ at one point each. It is this difference in risk profile, rather than uniform dominance, that motivates the preset.}
  \label{fig:rulecomp}
\end{figure}

Table~\ref{tab:points} lists the twenty parameter points with their asymptotic optima, rate-gain factors and convergence diagnostics. Tables~\ref{tab:regret_median} and \ref{tab:regret_max} give the calibration regret statistics of the four fixed rules, and Fig.~\ref{fig:rulecomp} displays the resulting risk profile.

The following counts, quoted in the main text, follow from the validated sweep outputs summarized in these tables. The multi-metric dominant band is contiguous at $20/20$ points and lies strictly between $\tau_{\mathrm{fid}}$ and $\tau_{C}$. The refined exact optimum satisfies $\tau_{\mathrm{finite}}(\epsilon) \le \tau_{C}$ at $20/20$ points, while strict monotonicity of $\tau_{\mathrm{finite}}$ with target stringency holds at only $5/20$ points and is therefore not claimed. Relative to $\tau_{C}$, the preset $\tau_{\mathrm{BR}}$ is superior at $20/20$ points simultaneously in overall median, worst-band median and maximum regret. Relative to $\tau_{\mathrm{fid}}$, $\tau_{\mathrm{BR}}$ lowers the worst-band median at $19/20$ points, whereas $\tau_{\mathrm{fid}}$ retains the lower overall median at $20/20$ points.

\begin{table*}[t]
\caption{Parameter points, asymptotic optima and convergence diagnostics. The standard-linewidth regime uses $\chi/2\pi = 1.2$~MHz and $\kappa/2\pi = 5.0$~MHz; the sub-MHz-linewidth regime uses $\chi/2\pi = 0.19$~MHz and $\kappa/2\pi = 0.80$~MHz, at the same ratio $\chi/\kappa = 0.24$. $G$ is the asymptotic rate-gain factor of Eq.~\eqref{eq:gain}; $100(G-1)$ is the corresponding percentage improvement. The interpolation error is the largest deviation, in dex of $\log_{10}(1/\epsilon)$, between the $\lambda_N(\tau)$ interpolant and independent exact evaluations at off-grid midpoints, measured before those midpoints are added as nodes. The grid error is the largest relative wall-clock difference between the tournament-grid optimum and the refined continuous-$\tau$ optimum.}
\label{tab:points}
\begin{ruledtabular}
\begin{tabular}{llrrrrrrrrrr}
Regime & Point & $T_1$ & $\eta$ & $\bar n$ & $\tau_{\mathrm{oh}}$ & $\tau_{\mathrm{fid}}$ & $\tau_{C}$ & $G$ & $100(G-1)$ & interp.\ err. & grid err. \\
 & & ($\mu$s) & & & ($\mu$s) & ($\mu$s) & ($\mu$s) & & (\%) & (dex) & (\%) \\
\colrule
Standard & P01\_baseline & 30 & 0.45 & 80 & 15 & 0.1415 & 0.2625 & 1.1436 & 14.36 & 0.0028 & 0.08 \\
Standard & P02\_lowEta & 30 & 0.30 & 80 & 15 & 0.1710 & 0.3177 & 1.1374 & 13.74 & 0.0020 & 0.10 \\
Standard & P03\_highEta & 30 & 0.65 & 80 & 15 & 0.1202 & 0.2231 & 1.1484 & 14.84 & 0.0036 & 0.07 \\
Standard & P04\_shortT1 & 20 & 0.45 & 80 & 15 & 0.1377 & 0.2508 & 1.1417 & 14.17 & 0.0023 & 0.08 \\
Standard & P05\_longT1 & 100 & 0.45 & 80 & 15 & 0.1519 & 0.2985 & 1.1516 & 15.16 & 0.0047 & 0.09 \\
Standard & P06\_lowNbar & 30 & 0.45 & 40 & 15 & 0.1974 & 0.3666 & 1.1316 & 13.16 & 0.0015 & 0.11 \\
Standard & P07\_highNbar & 20 & 0.50 & 120 & 30 & 0.1106 & 0.2047 & 1.1498 & 14.98 & 0.0034 & 0.03 \\
Standard & P08\_lowOh & 30 & 0.45 & 80 & 5 & 0.1415 & 0.2446 & 1.1276 & 12.76 & 0.0028 & 0.22 \\
Standard & P09\_highOh & 30 & 0.45 & 80 & 30 & 0.1415 & 0.2685 & 1.1482 & 14.82 & 0.0030 & 0.04 \\
Standard & P10\_bestCase & 140 & 0.60 & 80 & 15 & 0.1353 & 0.2689 & 1.1572 & 15.72 & 0.0065 & 0.09 \\
Sub-MHz & P01\_baseline & 30 & 0.45 & 80 & 15 & 0.7840 & 1.1994 & 1.1011 & 10.11 & 0.0014 & 0.34 \\
Sub-MHz & P02\_lowEta & 30 & 0.30 & 80 & 15 & 0.9318 & 1.4006 & 1.0925 & 9.25 & 0.0006 & 0.39 \\
Sub-MHz & P03\_highEta & 30 & 0.65 & 80 & 15 & 0.6760 & 1.0472 & 1.1063 & 10.63 & 0.0009 & 0.30 \\
Sub-MHz & P04\_shortT1 & 20 & 0.45 & 80 & 15 & 0.7604 & 1.1489 & 1.0986 & 9.86 & 0.0019 & 0.32 \\
Sub-MHz & P05\_longT1 & 100 & 0.45 & 80 & 15 & 0.8525 & 1.3455 & 1.1063 & 10.63 & 0.0037 & 0.39 \\
Sub-MHz & P06\_lowNbar & 30 & 0.45 & 40 & 15 & 1.0601 & 1.5696 & 1.0849 & 8.49 & 0.0014 & 0.43 \\
Sub-MHz & P07\_highNbar & 20 & 0.50 & 120 & 30 & 0.6248 & 0.9975 & 1.1145 & 11.45 & 0.0009 & 0.15 \\
Sub-MHz & P08\_lowOh & 30 & 0.45 & 80 & 5 & 0.7840 & 1.0586 & 1.0626 & 6.26 & 0.0016 & 0.80 \\
Sub-MHz & P09\_highOh & 30 & 0.45 & 80 & 30 & 0.7840 & 1.2636 & 1.1160 & 11.60 & 0.0007 & 0.19 \\
Sub-MHz & P10\_bestCase & 140 & 0.60 & 80 & 15 & 0.7688 & 1.2370 & 1.1132 & 11.32 & 0.0023 & 0.36 \\
\end{tabular}
\end{ruledtabular}
\end{table*}

\begin{table*}[t]
\caption{Overall median and worst-band median calibration regret (\%) of the four fixed integration-time rules. Regret is defined in Eq.~\eqref{eq:regret} and the summary statistics in Appendix~\ref{app:metrics}. The worst-band median is the largest median regret among the four bands $[3,5]$, $[5,7]$, $[7,9]$ and $[9,12]$ in $\log_{10}(1/\epsilon)$.}
\label{tab:regret_median}
\begin{ruledtabular}
\begin{tabular}{llrrrrrrrr}
Regime & Point & \multicolumn{2}{c}{$\tau_{\mathrm{fid}}$} & \multicolumn{2}{c}{$\tau_{\mathrm{BR}}$} & \multicolumn{2}{c}{$\tau_C$} & \multicolumn{2}{c}{best fixed} \\
\cline{3-10}
 & & med. & w.-band & med. & w.-band & med. & w.-band & med. & w.-band \\
\colrule
Standard & P01\_baseline & 0.21 & 19.91 & 0.61 & 0.78 & 0.90 & 1.06 & 0.21 & 19.91 \\
Standard & P02\_lowEta & 0.28 & 19.81 & 0.73 & 0.94 & 1.08 & 1.27 & 0.27 & 19.80 \\
Standard & P03\_highEta & 0.15 & 0.22 & 0.53 & 0.67 & 0.77 & 0.91 & 0.15 & 0.22 \\
Standard & P04\_shortT1 & 0.21 & 19.81 & 0.56 & 0.71 & 0.84 & 0.97 & 0.23 & 0.32 \\
Standard & P05\_longT1 & 0.22 & 24.74 & 0.77 & 0.94 & 1.11 & 1.28 & 0.25 & 0.37 \\
Standard & P06\_lowNbar & 0.33 & 19.60 & 0.83 & 1.06 & 1.24 & 1.44 & 0.32 & 19.57 \\
Standard & P07\_highNbar & 0.07 & 19.96 & 0.24 & 0.30 & 0.35 & 0.41 & 0.07 & 19.96 \\
Standard & P08\_lowOh & 0.61 & 19.74 & 1.58 & 2.10 & 2.30 & 2.79 & 0.63 & 0.99 \\
Standard & P09\_highOh & 0.10 & 19.96 & 0.32 & 0.41 & 0.47 & 0.55 & 0.10 & 19.96 \\
Standard & P10\_bestCase & 0.18 & 24.91 & 0.53 & 0.68 & 1.02 & 1.16 & 0.18 & 24.91 \\
Sub-MHz & P01\_baseline & 1.08 & 19.18 & 1.91 & 2.76 & 2.85 & 3.70 & 1.00 & 1.39 \\
Sub-MHz & P02\_lowEta & 1.39 & 13.45 & 2.44 & 3.42 & 3.09 & 4.09 & 1.51 & 2.23 \\
Sub-MHz & P03\_highEta & 0.84 & 15.55 & 1.78 & 2.52 & 2.63 & 3.36 & 0.85 & 1.30 \\
Sub-MHz & P04\_shortT1 & 1.06 & 13.48 & 1.74 & 2.47 & 2.61 & 3.35 & 0.91 & 1.16 \\
Sub-MHz & P05\_longT1 & 1.11 & 18.08 & 2.43 & 3.28 & 3.55 & 4.42 & 1.11 & 18.08 \\
Sub-MHz & P06\_lowNbar & 1.82 & 12.68 & 2.60 & 3.64 & 3.27 & 4.37 & 1.61 & 2.32 \\
Sub-MHz & P07\_highNbar & 0.41 & 19.49 & 0.90 & 1.22 & 1.34 & 1.65 & 0.41 & 0.62 \\
Sub-MHz & P08\_lowOh & 3.02 & 17.80 & 4.42 & 6.77 & 5.28 & 7.73 & 2.87 & 4.38 \\
Sub-MHz & P09\_highOh & 0.55 & 19.58 & 1.12 & 1.55 & 1.68 & 2.10 & 0.53 & 0.79 \\
Sub-MHz & P10\_bestCase & 1.05 & 19.49 & 2.40 & 3.29 & 3.46 & 4.32 & 1.04 & 19.24 \\
\end{tabular}
\end{ruledtabular}
\end{table*}

\begin{table*}[t]
\caption{Maximum calibration regret (\%) and the fraction (\%) of the target interval on which each rule exceeds $5\%$ regret. Regret is defined in Eq.~\eqref{eq:regret} and the summary statistics in Appendix~\ref{app:metrics}. Maxima are taken as the larger of the maximum on the sampled target grid over the declared interval and the exact right-hand-limit values at the integer-shot breakpoints of each rule.}
\label{tab:regret_max}
\begin{ruledtabular}
\begin{tabular}{llrrrrrrrr}
Regime & Point & \multicolumn{2}{c}{$\tau_{\mathrm{fid}}$} & \multicolumn{2}{c}{$\tau_{\mathrm{BR}}$} & \multicolumn{2}{c}{$\tau_C$} & \multicolumn{2}{c}{best fixed} \\
\cline{3-10}
 & & max. & $>5\%$ & max. & $>5\%$ & max. & $>5\%$ & max. & $>5\%$ \\
\colrule
Standard & P01\_baseline & 50.0 & 29.2 & 101.1 & 2.3 & 101.7 & 5.4 & 50.0 & 29.2 \\
Standard & P02\_lowEta & 50.0 & 34.3 & 50.9 & 2.2 & 51.4 & 5.4 & 50.0 & 27.9 \\
Standard & P03\_highEta & 50.0 & 24.6 & 101.1 & 2.6 & 101.6 & 5.8 & 50.0 & 24.6 \\
Standard & P04\_shortT1 & 50.0 & 33.7 & 50.7 & 2.2 & 51.1 & 6.7 & 50.0 & 16.2 \\
Standard & P05\_longT1 & 50.0 & 26.4 & 101.5 & 2.8 & 102.2 & 5.2 & 100.2 & 15.8 \\
Standard & P06\_lowNbar & 50.0 & 34.9 & 51.0 & 2.0 & 51.7 & 6.8 & 50.0 & 29.1 \\
Standard & P07\_highNbar & 50.0 & 27.4 & 50.3 & 1.9 & 50.5 & 4.6 & 50.0 & 27.4 \\
Standard & P08\_lowOh & 50.0 & 29.2 & 102.8 & 1.4 & 104.2 & 4.0 & 100.6 & 15.3 \\
Standard & P09\_highOh & 50.0 & 29.2 & 100.6 & 2.6 & 100.9 & 5.8 & 50.0 & 29.2 \\
Standard & P10\_bestCase & 50.0 & 19.5 & 101.0 & 1.5 & 102.0 & 5.2 & 50.0 & 19.5 \\
Sub-MHz & P01\_baseline & 50.0 & 40.4 & 52.2 & 0.8 & 53.9 & 2.6 & 50.0 & 23.8 \\
Sub-MHz & P02\_lowEta & 50.0 & 43.8 & 53.1 & 1.6 & 54.3 & 1.9 & 50.6 & 12.2 \\
Sub-MHz & P03\_highEta & 50.0 & 34.6 & 52.1 & 0.6 & 53.5 & 3.0 & 50.0 & 18.1 \\
Sub-MHz & P04\_shortT1 & 50.0 & 42.9 & 52.1 & 1.6 & 53.7 & 2.6 & 50.0 & 25.9 \\
Sub-MHz & P05\_longT1 & 50.0 & 31.3 & 52.6 & 0.7 & 54.4 & 6.8 & 50.0 & 31.3 \\
Sub-MHz & P06\_lowNbar & 50.0 & 47.3 & 53.2 & 2.3 & 54.6 & 4.6 & 33.3 & 13.6 \\
Sub-MHz & P07\_highNbar & 50.0 & 36.6 & 51.1 & 0.8 & 51.9 & 3.8 & 33.3 & 15.5 \\
Sub-MHz & P08\_lowOh & 50.0 & 40.4 & 54.4 & 38.5 & 56.2 & 55.6 & 50.0 & 23.7 \\
Sub-MHz & P09\_highOh & 50.0 & 40.4 & 51.4 & 0.9 & 52.4 & 4.1 & 50.0 & 17.7 \\
Sub-MHz & P10\_bestCase & 50.0 & 29.1 & 52.5 & 0.7 & 54.3 & 3.8 & 50.0 & 22.4 \\
\end{tabular}
\end{ruledtabular}
\end{table*}

\section{Bahadur--Rao design-target selection and prefactor validation}

The prefactor $A(\tau)$ of Eq.~\eqref{eq:bahadur_rao} in the main text is extracted from the plateau of
\begin{equation}
  g(N;\tau) = \log P_{\mathrm{e}}^{(N)}(\tau) + N\,C(\tau)
              + \tfrac{1}{2}\log N
  \label{eq:plateau}
\end{equation}
evaluated at $N \in \{26,30,34,38\}$. A given $\tau$ enters the interpolation for $A$ only if the spread of $g$ over those shot counts is below $0.05$; $A(\tau)$ is then the exponential of the mean of $g$.

\begin{figure*}[t]
  \centering
  \includegraphics[width=\textwidth]{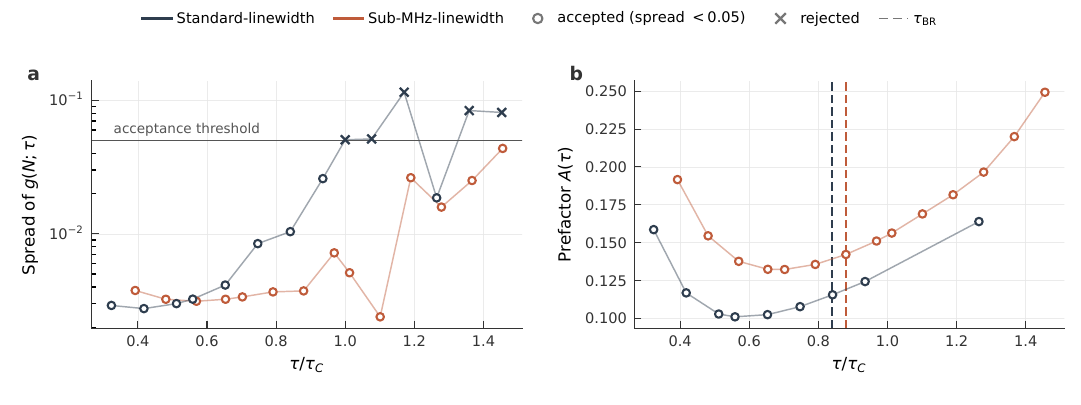}
  \caption{\textbf{Bahadur--Rao prefactor validation at both baselines.} (a)~Spread of $g(N;\tau)$ of Eq.~\eqref{eq:plateau} over $N \in \{26,30,34,38\}$, against $\tau/\tau_{C}$; the horizontal line marks the acceptance threshold. (b)~Extracted prefactor $A(\tau)$ over the accepted set. Rejected points are excluded from the interpolation used to evaluate $A$ at the selected preset.}
  \label{fig:plateau}
\end{figure*}

Figure~\ref{fig:plateau} shows the diagnostic at both baselines, with $\tau$ normalised to $\tau_{C}$ so the two regimes are directly comparable. At the sub-MHz baseline all $14$ sampled $\tau$ satisfy the criterion, with spreads between $0.0024$ and $0.044$. At the standard-linewidth baseline $9$ of $14$ sampled values satisfy the criterion. The five rejected samples occur among the longer integration windows and have spreads up to $0.115$, indicating that the sampled shot counts at those points have not yet reached the sharp large-deviation regime. The accepted set is not contiguous: the eight shortest sampled windows ($\tau\le0.242~\mu$s) are accepted, the next three are rejected, one isolated longer window at $\tau=0.332~\mu$s is accepted, and the two longest are rejected. The selected preset $\tau_{\mathrm{BR}} = 0.221~\mu$s lies inside the contiguous lower accepted block, so the prefactor used in the selection is interpolated locally in $\tau$ rather than extrapolated. This concerns the integration-time coordinate only: at $\epsilon_{\mathrm{design}}=10^{-4}$ the surrogate objective is minimised at a continuous shot count $N_{\mathrm{BR}}(\tau_{\mathrm{BR}})\approx1.3$, well below the range $N\in\{26,\dots,38\}$ at which $A(\tau)$ was validated. The Bahadur--Rao form is therefore used only to select a candidate preset; every regret statistic reported anywhere in this work is computed from exact $P_{\mathrm{e}}^{(N)}$ tables, never from the Bahadur--Rao form itself. We note that $\tau_{C}=0.263~\mu$s is itself among the rejected points at that baseline.

Table~\ref{tab:brselect} lists, for each candidate design target, the selected preset and its regret statistics. When multiple candidates attain the same criterion value, ties are resolved in favor of the shallowest design target. With this convention, both the minimax and the minimum-median criterion select $\epsilon_{\mathrm{design}} = 10^{-4}$ at all twenty points, which fixes the prescription adopted in the main text. Each criterion selects $\epsilon_{\mathrm{design}} = 10^{-4}$ strictly at eighteen of the twenty points, the remaining two being exact ties resolved by that convention. At ten of the twenty points the selections for $\epsilon_{\mathrm{design}} = 10^{-8}$ and $10^{-12}$ coincide, because both fall on the same node of the selection grid; the corresponding column pairs of Table~\ref{tab:brselect} are then identical.

\begin{table*}[t]
\caption{Bahadur--Rao design-target selection. For each candidate design target $\epsilon_{\mathrm{design}}$ the table lists the selected preset $\tau_{\mathrm{BR}}$, in $\mu$s, and its overall median and maximum calibration regret over $\log_{10}(1/\epsilon)\in[3,12]$. When criterion values tie, the shallowest design target is selected.}
\label{tab:brselect}
\begin{ruledtabular}
\begin{tabular}{llrrrrrrrrr}
Regime & Point & \multicolumn{3}{c}{$\epsilon_{\mathrm{design}}=10^{-4}$} & \multicolumn{3}{c}{$\epsilon_{\mathrm{design}}=10^{-8}$} & \multicolumn{3}{c}{$\epsilon_{\mathrm{design}}=10^{-12}$} \\
\cline{3-11}
 & & $\tau_{\mathrm{BR}}$ & med. & max. & $\tau_{\mathrm{BR}}$ & med. & max. & $\tau_{\mathrm{BR}}$ & med. & max. \\
\colrule
Standard & P01\_baseline & 0.2205 & 0.61 & 101.1 & 0.2335 & 0.70 & 101.3 & 0.2453 & 0.78 & 101.5 \\
Standard & P02\_lowEta & 0.2676 & 0.73 & 50.9 & 0.2964 & 0.93 & 51.2 & 0.2973 & 0.94 & 51.2 \\
Standard & P03\_highEta & 0.1872 & 0.53 & 101.1 & 0.1981 & 0.60 & 101.2 & 0.1981 & 0.60 & 101.2 \\
Standard & P04\_shortT1 & 0.2115 & 0.56 & 50.7 & 0.2239 & 0.65 & 50.8 & 0.2350 & 0.73 & 50.9 \\
Standard & P05\_longT1 & 0.2481 & 0.77 & 101.5 & 0.2481 & 0.77 & 101.5 & 0.2481 & 0.77 & 101.5 \\
Standard & P06\_lowNbar & 0.3084 & 0.83 & 51.0 & 0.3427 & 1.07 & 51.4 & 0.3427 & 1.07 & 51.4 \\
Standard & P07\_highNbar & 0.1718 & 0.24 & 50.3 & 0.1724 & 0.24 & 50.3 & 0.1822 & 0.28 & 50.3 \\
Standard & P08\_lowOh & 0.2093 & 1.58 & 102.8 & 0.2311 & 2.02 & 103.7 & 0.2317 & 2.04 & 103.7 \\
Standard & P09\_highOh & 0.2245 & 0.32 & 100.6 & 0.2378 & 0.37 & 100.7 & 0.2499 & 0.41 & 100.8 \\
Standard & P10\_bestCase & 0.1971 & 0.53 & 101.0 & 0.1971 & 0.53 & 101.0 & 0.1971 & 0.53 & 101.0 \\
Sub-MHz & P01\_baseline & 1.0554 & 1.91 & 52.2 & 1.1609 & 2.59 & 53.5 & 1.1609 & 2.59 & 53.5 \\
Sub-MHz & P02\_lowEta & 1.2973 & 2.44 & 53.1 & 1.3604 & 2.83 & 53.8 & 1.3604 & 2.83 & 53.8 \\
Sub-MHz & P03\_highEta & 0.9186 & 1.78 & 52.1 & 0.9663 & 2.09 & 52.6 & 1.0111 & 2.39 & 53.1 \\
Sub-MHz & P04\_shortT1 & 1.0142 & 1.74 & 52.1 & 1.1148 & 2.38 & 53.3 & 1.1148 & 2.38 & 53.3 \\
Sub-MHz & P05\_longT1 & 1.1749 & 2.43 & 52.6 & 1.2946 & 3.21 & 53.9 & 1.2946 & 3.21 & 53.9 \\
Sub-MHz & P06\_lowNbar & 1.4588 & 2.60 & 53.2 & 1.5290 & 3.02 & 54.1 & 1.5290 & 3.02 & 54.1 \\
Sub-MHz & P07\_highNbar & 0.8687 & 0.90 & 51.1 & 0.9145 & 1.06 & 51.4 & 0.9577 & 1.20 & 51.6 \\
Sub-MHz & P08\_lowOh & 1.0055 & 4.42 & 54.4 & 1.0082 & 4.46 & 54.5 & 1.0512 & 5.15 & 55.9 \\
Sub-MHz & P09\_highOh & 1.0978 & 1.12 & 51.4 & 1.2110 & 1.50 & 52.1 & 1.2110 & 1.50 & 52.1 \\
Sub-MHz & P10\_bestCase & 1.0763 & 2.40 & 52.5 & 1.1839 & 3.11 & 53.7 & 1.1870 & 3.13 & 53.7 \\
\end{tabular}
\end{ruledtabular}
\end{table*}

\section{Fast-reset limit}
\label{sec:fast_reset}

The main-text survey uses $\tau_{\mathrm{oh}}=5$--$30~\mu$s. Because active reset can reach the few-hundred-nanosecond scale \cite{Magnard2018fast,krinner2022realizing}, we repeat the asymptotic optimization at the standard-linewidth baseline down to $\tau_{\mathrm{oh}}=0$. Table~\ref{tab:fastreset} lists the resulting optima; its entries at $\tau_{\mathrm{oh}}=5$, $15$ and $30~\mu$s reproduce the corresponding rows of Table~\ref{tab:points} exactly, since both are produced by the same pipeline.

The interior maximum of $\mathcal{R}_C(\tau)=C(\tau)/(\tau+\tau_{\mathrm{oh}})$ remains unique at every overhead tested, including $\tau_{\mathrm{oh}}=0$, where the stationarity condition reduces to $C(\tau)=\tau C'(\tau)$ and is satisfied because $T_1$ smearing saturates $C(\tau)$. The framework therefore remains well posed in the fast-reset limit.

Since $\tau_{\mathrm{fid}}$ is overhead-independent while $\mathrm{d}\tau_C/\mathrm{d}\tau_{\mathrm{oh}}>0$, the two optima cross at a single overhead, near $\tau_{\mathrm{oh}}=0.10~\mu$s at this baseline, where $G=1$. The rate gain is therefore not monotonic in $\tau_{\mathrm{oh}}$: it falls to unity at the crossing and rises again on both sides. Below the crossing the ordering reverses and $\tau_C<\tau_{\mathrm{fid}}$: with no overhead to amortize, the rate objective favours a shorter window than the fidelity objective. For total per-shot overheads in the range $\tau_{\mathrm{oh}}=0.2$--$0.5~\mu\mathrm{s}$, $\tau_C$ remains longer than $\tau_{\mathrm{fid}}$, but the asymptotic rate gain is only $0.8\%$--$4.0\%$. A sub-microsecond reset operation does not by itself imply a sub-microsecond value of $\tau_{\mathrm{oh}}$, because the latter also includes resonator depletion, classical processing, and sequence re-arming.

\begin{table}[t]
\caption{Asymptotic optima at the standard-linewidth baseline ($T_1=30~\mu$s, $\eta=0.45$, $\bar n=80$) in the fast-reset regime, together with selected reference points from the main-text overhead range. $\tau_{\mathrm{fid}}$ is independent of $\tau_{\mathrm{oh}}$. The last column counts interior maxima of $\mathcal{R}_C(\tau)$ on $\tau\in[0.01,0.80]~\mu$s.}
\label{tab:fastreset}
\begin{ruledtabular}
\begin{tabular}{rrrrr}
$\tau_{\mathrm{oh}}$ ($\mu$s) & $\tau_{C}$ ($\mu$s) &
$\tau_{C}/\tau_{\mathrm{fid}}$ & $G$ & maxima \\
\colrule
 0.00 & 0.1170 & 0.8266 & 1.0427 & 1 \\
 0.10 & 0.1414 & 0.9994 & 1.0000 & 1 \\
 0.20 & 0.1551 & 1.0959 & 1.0076 & 1 \\
 0.30 & 0.1649 & 1.1654 & 1.0193 & 1 \\
 0.50 & 0.1789 & 1.2642 & 1.0400 & 1 \\
 1.00 & 0.1998 & 1.4123 & 1.0720 & 1 \\
 2.00 & 0.2209 & 1.5614 & 1.1010 & 1 \\
 5.00 & 0.2446 & 1.7290 & 1.1276 & 1 \\
15.00 & 0.2625 & 1.8556 & 1.1436 & 1 \\
30.00 & 0.2685 & 1.8977 & 1.1482 & 1 \\
\end{tabular}
\end{ruledtabular}
\end{table}

\section{Driven-cavity response and the reduced model}
\label{app:response}

Equation~\eqref{eq:branch} is the response model used throughout the main text and in the entire finite-shot study. This appendix gives the exact response for the readout protocol assumed here, identifies the approximations that separate the two, and quantifies the consequences for the asymptotic optima.

\paragraph{Protocol and exact solution.}
The statements below are specific to the following protocol: (i) the cavity is empty at $t=0$; (ii) the readout pulse has an abrupt, constant envelope of amplitude $\varepsilon$; (iii) the drive is placed at the bare-cavity frequency, midway between the two dispersively shifted resonances; and (iv) homodyne detection is performed on the quadrature along which the two qubit-conditioned branches separate. Under (i)--(iii) the qubit-conditioned intracavity amplitudes obey
\begin{equation}
\dot{\alpha}_{\pm}(t) = -\left(\frac{\kappa}{2} \pm i\chi\right)\alpha_{\pm}(t) - i\varepsilon ,
\qquad \alpha_{\pm}(0)=0 ,
\end{equation}
with solution
\begin{equation}
\alpha_{\pm}(t) = \frac{-i\varepsilon}{\kappa/2 \pm i\chi}
\left[1 - e^{-(\kappa/2 \pm i\chi)t}\right].
\end{equation}
For the symmetric resonant-drive convention considered here, the qubit-conditioned cavity-state separation lies along a fixed measurement quadrature, so the relevant separation can be represented by a real scalar response. Projecting the physical pointer-state separation onto that quadrature at the same steady-state photon number $\bar n = |\alpha_{\pm}(\infty)|^{2} = |\varepsilon|^{2}/[(\kappa/2)^{2}+\chi^{2}]$, and normalizing the resulting response so that it enters the SNR and matched-filter definitions in the same convention as Eq.~\eqref{eq:branch}, gives
\begin{equation}
\Delta\alpha_{\mathrm{exact}}(t) = \frac{4\chi}{\kappa}\,
\frac{1 - e^{-\kappa t/2}\!\left[\cos\chi t + (\kappa/2\chi)\sin\chi t\right]}
{\sqrt{1+(2\chi/\kappa)^{2}}} .
\label{eq:dalpha_exact}
\end{equation}
At fixed $\bar n$ the steady-state separation is therefore smaller than that of Eq.~\eqref{eq:branch} by the factor $[1+(2\chi/\kappa)^{2}]^{-1/2}$, which equals $0.902$ at $\chi/\kappa = 0.24$; this is the familiar reduction of the branch separation by the dispersive detuning of each branch from the drive.

\paragraph{Relation to the reduced model.}
Equation~\eqref{eq:dalpha_exact} is not reduced to Eq.~\eqref{eq:branch} by any choice of $\chi/\kappa$ under this protocol. Equation~\eqref{eq:branch} describes a single-pole amplitude build-up and is linear in $t$ at short times, whereas the exact separation requires both intracavity field build-up and the accumulation of the dispersive phase, and is therefore quadratic in $t$: expanding Eq.~\eqref{eq:dalpha_exact} gives $\Delta\alpha_{\mathrm{exact}}(t) \simeq (4\chi/\kappa)(\kappa^{2}/4+\chi^{2})t^{2}/2$ up to the normalization factor. In the limit $\chi \to 0$ the bracket in Eq.~\eqref{eq:dalpha_exact} becomes the double-pole form $1-(1+\kappa t/2)e^{-\kappa t/2}$ rather than the single-pole form $1-e^{-\kappa t/2}$ of Eq.~\eqref{eq:branch}, so the difference in normalized temporal shape does not vanish in the weak-dispersive limit, although both separations themselves vanish there with the common factor $4\chi/\kappa$. This short-time difference is the origin of the exponents quoted in the caption of Fig.~\ref{fig:physical_model}: Eq.~\eqref{eq:branch} gives $SNR^{2}\propto\tau^{3}$, Eq.~\eqref{eq:dalpha_exact} gives $SNR^{2}\propto\tau^{5}$. For the initially empty, qubit-independent dispersive drive considered here, the branch separation has zero initial slope, $\dot{\Delta}(0)\propto\chi[\alpha_{+}(0)+\alpha_{-}(0)]=0$. Equation~\eqref{eq:branch} is therefore used as a phenomenological single-pole response rather than as the exact transient of this protocol.

\paragraph{Effect on the asymptotic optima.}
Table~\ref{tab:responsecomp} repeats the asymptotic optimization at all ten standard-linewidth survey points with Eq.~\eqref{eq:dalpha_exact} in place of Eq.~\eqref{eq:branch}, the post-jump differential evolution taken as $e^{-(\kappa/2-i\chi)(t-t_j)}$ in the same quadrature, and all other model elements unchanged. Both response models give $\tau_C > \tau_{\mathrm{fid}}$ at $10/10$ points. The exact response shifts both optima to longer times, leaves their ratio close to that of the reduced model ($\tau_C/\tau_{\mathrm{fid}} = 1.66$--$1.93$, against $1.73$--$1.99$ in Table~\ref{tab:points}), and slightly increases the asymptotic rate gain, to $14.3\%$--$17.6\%$ against $12.8\%$--$15.7\%$. At the standard-linewidth baseline the separation becomes $\tau_C/\tau_{\mathrm{fid}} = 1.81$, i.e.\ approximately $81\%$, against $86\%$ for Eq.~\eqref{eq:branch}. The qualitative conclusions of the paper --- the existence of the separation, its sign, its order of magnitude, and its dependence on $\tau_{\mathrm{oh}}$ --- are therefore unchanged. We emphasize that the full finite-shot, regret and Bahadur--Rao analyses reported in the main text are computed with Eq.~\eqref{eq:branch} and are \emph{not} re-validated against Eq.~\eqref{eq:dalpha_exact} here; the comparison in Table~\ref{tab:responsecomp} is confined to the asymptotic optima. All numerical values quoted in the main text and in Tables~\ref{tab:points}--\ref{tab:fastreset} refer to Eq.~\eqref{eq:branch}, and the two sets of values are not interchangeable.

\begin{table}[t]
\caption{Asymptotic optima under the exact driven-cavity response of Eq.~\eqref{eq:dalpha_exact}, at the ten standard-linewidth survey points of Table~\ref{tab:points}. All other model elements, including $T_1$, $\eta$, $\bar n$ and $\tau_{\mathrm{oh}}$, are identical to Table~\ref{tab:points}. The two rightmost columns compare the resulting separation with the corresponding ratio formed from the Eq.~\eqref{eq:branch} entries of Table~\ref{tab:points}. Integration times are in $\mu$s.}
\label{tab:responsecomp}
\begin{ruledtabular}
\begin{tabular}{lrrrrr}
 & \multicolumn{3}{c}{Eq.~\eqref{eq:dalpha_exact}} & \multicolumn{2}{c}{$\tau_{C}/\tau_{\mathrm{fid}}$} \\
Point & $\tau_{\mathrm{fid}}$ & $\tau_{C}$ & $100(G-1)$ & exact & Eq.~\eqref{eq:branch} \\
 & ($\mu$s) & ($\mu$s) & (\%) & & \\
\colrule
P01\_baseline & 0.2001 & 0.3619 & 16.35 & 1.809 & 1.855 \\
P02\_lowEta   & 0.2340 & 0.4289 & 15.38 & 1.833 & 1.858 \\
P03\_highEta  & 0.1755 & 0.3139 & 17.13 & 1.789 & 1.856 \\
P04\_shortT1  & 0.1955 & 0.3461 & 16.18 & 1.770 & 1.821 \\
P05\_longT1   & 0.2130 & 0.4102 & 16.85 & 1.926 & 1.965 \\
P06\_lowNbar  & 0.2640 & 0.4879 & 14.62 & 1.848 & 1.857 \\
P07\_highNbar & 0.1640 & 0.2923 & 17.64 & 1.782 & 1.851 \\
P08\_lowOh    & 0.2001 & 0.3320 & 14.26 & 1.659 & 1.729 \\
P09\_highOh   & 0.2001 & 0.3730 & 16.98 & 1.864 & 1.898 \\
P10\_bestCase & 0.1940 & 0.3746 & 17.60 & 1.931 & 1.988 \\
\end{tabular}
\end{ruledtabular}
\end{table}

\begin{table}[t]
\caption{Integration time selected by the empirically minimum-median rule
(``best fixed'' in Tables~\ref{tab:regret_median} and \ref{tab:regret_max}) at each
parameter point, with $\tau_{\mathrm{fid}}$ repeated from Table~\ref{tab:points} for
comparison. Both times are in $\mu$s. Selection is performed on the coarse candidate
grid described in Appendix~\ref{app:metrics}; the values are those of the validated
sweep used throughout.}
\label{tab:bestfixed}
\begin{ruledtabular}
\begin{tabular}{llrr}
Regime & Point & $\tau_{\mathrm{fid}}$ ($\mu$s) & best-fixed $\tau$ ($\mu$s) \\
\colrule
Standard & P01\_baseline & 0.1415 & 0.1415 \\
Standard & P02\_lowEta & 0.1710 & 0.1774 \\
Standard & P03\_highEta & 0.1202 & 0.1202 \\
Standard & P04\_shortT1 & 0.1377 & 0.1531 \\
Standard & P05\_longT1 & 0.1519 & 0.1625 \\
Standard & P06\_lowNbar & 0.1974 & 0.2047 \\
Standard & P07\_highNbar & 0.1106 & 0.1106 \\
Standard & P08\_lowOh & 0.1415 & 0.1526 \\
Standard & P09\_highOh & 0.1415 & 0.1415 \\
Standard & P10\_bestCase & 0.1353 & 0.1353 \\
Sub-MHz & P01\_baseline & 0.7840 & 0.8424 \\
Sub-MHz & P02\_lowEta & 0.9318 & 1.1142 \\
Sub-MHz & P03\_highEta & 0.6760 & 0.7318 \\
Sub-MHz & P04\_shortT1 & 0.7604 & 0.8110 \\
Sub-MHz & P05\_longT1 & 0.8525 & 0.8525 \\
Sub-MHz & P06\_lowNbar & 1.0601 & 1.2546 \\
Sub-MHz & P07\_highNbar & 0.6248 & 0.6889 \\
Sub-MHz & P08\_lowOh & 0.7840 & 0.8727 \\
Sub-MHz & P09\_highOh & 0.7840 & 0.8694 \\
Sub-MHz & P10\_bestCase & 0.7688 & 0.7959 \\
\end{tabular}
\end{ruledtabular}
\end{table}

\paragraph{Binarized readout.}
The rate advantage quantified by Eq.~\eqref{eq:gain} is obtained from the full analog score distributions. For a binarized acquisition chain, the corresponding optimization must instead be performed using the Chernoff information of the induced binary channel, which depends on the assignment rule; we do not pursue that separate optimization here.

\section{Exact finite-shot engine: implementation, convergence and interpolation tests}

\paragraph{Exact $N$-shot error probabilities.}
The per-shot log-likelihood ratio $L(x) = \log[p_e(x)/p_g(x)]$ is binned onto a $16384$-point grid by linear mass splitting under each hypothesis. Each binned distribution is exponentially tilted so that the tilted mean vanishes, raised to the $N$th convolution power by a single forward FFT and $N$ inverse transforms, and the tilt is then unwound. This keeps the far tail numerically stable, which a direct convolution does not. The engine is validated against the closed-form Gaussian result $Q(\sqrt{N}\,\mathrm{SNR}/2)$ to within $6\times10^{-5}$ relative down to $P_{\mathrm{e}} \sim 2\times10^{-54}$.

\paragraph{Continuous-$\tau$ optima.}
For each shot count the threshold curve $\lambda_N(\tau) = -\log_{10}P_{\mathrm{e}}^{(N)}(\tau)$ is interpolated on its monotonically increasing branch, from the shortest sampled integration time up to the maximum of $\lambda_N$, by a monotone (PCHIP) interpolant, which is then inverted to give $\tau_N(\lambda)$. The refined optimum is $T_{\mathrm{ref}}(\lambda) = \min_N N[\tau_N(\lambda) + \tau_{\mathrm{oh}}]$. Interpolation accuracy is measured, not assumed: independent exact evaluations of $P_{\mathrm{e}}^{(N)}$ at off-grid midpoints, performed \emph{before} those midpoints are added as interpolation nodes, bound the error at $\le 0.0065$~dex across all twenty points (Table~\ref{tab:points}). The midpoints are then added, so the interpolants actually used are better than this bound.

\paragraph{Grid versus refined optimum.}
Comparing the tournament-grid optimum with the refined continuous-$\tau$ optimum quantifies the error that a grid-only treatment would incur: at most $0.80\%$ in wall-clock time at the worst point, with a median across points of $0.17\%$.

\paragraph{Integer-shot transitions.}
Transitions of every analysed rule are resolved as exact breakpoints $\lambda_N = -\log_{10}P_{\mathrm{e}}^{(N)}(\tau_{\mathrm{rule}})$ obtained from a dedicated exact table at that $\tau$, rather than by sampling. Maximum regret is taken as the larger of the maximum over the sampled target grid on the declared interval and the exact right-hand-limit values at those breakpoints, and the edges of the $>5\%$ windows are refined by bisection in $\log_{10}(1/\epsilon)$. The window widths quoted in the main text exceed the interpolation-limited edge uncertainty at the corresponding parameter point by more than an order of magnitude.

\section*{Data Availability}
The numerical datasets and scripts used to generate the figures in this study are available at Zenodo: \url{https://doi.org/10.5281/zenodo.21764035}.

\bibliography{references}

@article{danjou2021generalized,
  title={Generalized figure of merit for qubit readout},
  author={D'Anjou, Benjamin},
  journal={Physical Review A},
  volume={103},
  number={4},
  pages={042404},
  year={2021},
  doi={10.1103/PhysRevA.103.042404},
  publisher={APS}
}

@article{blais2020quantum,
  title={Quantum information processing and quantum optics with circuit quantum electrodynamics},
  author={Blais, Alexandre and Girvin, Steven M and Oliver, William D},
  journal={Nature Physics},
  volume={16},
  number={3},
  pages={247--256},
  year={2020},
  publisher={Nature Publishing Group UK London},
  doi = {10.1038/s41567-020-0806-z}
}

@article{walter2017rapid,
  title = {Rapid High-Fidelity Single-Shot Dispersive Readout of Superconducting Qubits},
  author = {Walter, T. and Kurpiers, P. and Gasparinetti, S. and Magnard, P. and Poto{\v{c}}nik, A. and Salath{\'{e}}, Y. and Pechal, M. and Mondal, M. and Oppliger, M. and Eichler, C. and Wallraff, A.},
  journal = {Physical Review Applied},
  volume = {7},
  issue = {5},
  pages = {054020},
  year = {2017},
  month = {May},
  doi = {10.1103/PhysRevApplied.7.054020},
  publisher = {American Physical Society}
}

@article{clerk2010introduction,
  title={Introduction to quantum noise, measurement, and amplification},
  author={Clerk, Aashish A and Devoret, Michel H and Girvin, Steven M and Marquardt, Florian and Schoelkopf, Robert J},
  journal={Reviews of Modern Physics},
  volume={82},
  number={2},
  pages={1155--1208},
  year={2010},
  publisher={APS},
  doi = {10.1103/RevModPhys.82.1155}
}

@article{gambetta2008quantum,
  title={Quantum trajectory approach to circuit QED: Quantum jumps and the Zeno effect},
  author={Gambetta, Jay and Blais, Alexandre and Boissonneault, Maxime and Houck, Andrew A and Schuster, DI and Girvin, Steven M},
  journal = {Phys. Rev. A},
  volume={77},
  number={1},
  pages={012112},
  year={2008},
  publisher={APS},
  doi = {10.1103/PhysRevA.77.012112}
}

@article{hatridge2013quantum,
  title={Quantum back-action of an individual variable-strength measurement},
  author={Hatridge, Michael and Shankar, Shyam and Mirrahimi, Mazyar and Schackert, F and Geerlings, K and Brecht, T and Sliwa, KM and Abdo, B and Frunzio, Luigi and Girvin, Steven M and others},
  journal={Science},
  volume={339},
  number={6116},
  pages={178--181},
  year={2013},
  doi={10.1126/science.1226897},
  publisher={American Association for the Advancement of Science}
}

@article{murch2013observing,
  title={Observing single quantum trajectories of a superconducting quantum bit},
  author={Murch, Kater W and Weber, SJ and Macklin, Christopher and Siddiqi, Irfan},
  journal={Nature},
  volume={502},
  number={7470},
  pages={211--214},
  year={2013},
  doi={10.1038/nature12539},
  publisher={Nature Publishing Group UK London}
}

@article{reed2010high,
  title={High-Fidelity Readout in Circuit Quantum Electrodynamics Using the Jaynes-Cummings Nonlinearity},
  author={Reed, Matthew D and DiCarlo, L and Johnson, BR and Sun, L and Schuster, DI and Frunzio, L and Schoelkopf, RJ},
  journal={Physical review letters},
  volume={105},
  number={17},
  pages={173601},
  year={2010},
  doi={10.1103/PhysRevLett.105.173601},
  publisher={APS}
}

@article{fechant2025offset,
  title={Offset charge dependence of measurement-induced transitions in transmons},
  author={F{\'e}chant, Mathieu and Dumas, Marie Fr{\'e}d{\'e}rique and B{\'e}n{\^a}tre, Denis and Gosling, Nicolas and Lenhard, Philipp and Spiecker, Martin and Geisert, Simon and Ihssen, S{\"o}ren and Wernsdorfer, Wolfgang and D’Anjou, Benjamin and others},
  journal={Physical Review Letters},
  volume={135},
  number={18},
  pages={180603},
  year={2025},
  doi={10.1103/yljv-b4kj},
  publisher={APS}
}

@article{jeffrey2014fast,
  title={Fast accurate state measurement with superconducting qubits},
  author={Jeffrey, Evan and Sank, Daniel and Mutus, JY and White, TC Friend and Kelly, J and Barends, R and Chen, Y and Chen, Z and Chiaro, B and Dunsworth, A and others},
  journal={Physical review letters},
  volume={112},
  number={19},
  pages={190504},
  year={2014},
  doi={10.1103/PhysRevLett.112.190504},
  publisher={APS}
}

@article{bugu2025entanglement,
  title={Resilience of Entanglement-Induced Coordination in Adversarial Environments: The Team-Based Quantum Sabotage Game},
  author={Bugu, Sinan},
  journal={arXiv preprint arXiv:2510.22444},
  year={2025}
}

@misc{bugu2026hfc,
      title={Hidden-Field Coordination Reveals Payoff-Free Quantum Correlation Structure in Decentralized Coordination}, 
      author={Sinan Bugu},
      year={2026},
      eprint={2601.21139},
      archivePrefix={arXiv},
      primaryClass={quant-ph},
      url={https://arxiv.org/abs/2601.21139}, 
}

@article{Chernoff1952,
  title={A measure of asymptotic efficiency for tests of a hypothesis based on the sum of observations},
  author={Chernoff, Herman},
  journal={The Annals of Mathematical Statistics},
  volume={23},
  number={4},
  pages={493--507},
  year={1952},
  doi = {10.1214/aoms/1177729330},
  publisher={Institute of Mathematical Statistics}
}

@book{CoverThomas2006,
  author = {Cover, Thomas M. and Thomas, Joy A.},
  title = {Elements of Information Theory},
  publisher = {Wiley-Interscience},
  year = {2006},
  doi = {10.1002/047174882X}
}

@article{pirandola2008computable,
  title={Computable bounds for the discrimination of Gaussian states},
  author={Pirandola, Stefano and Lloyd, Seth},
  journal={Physical Review A},
  volume={78},
  number={1},
  pages={012331},
  year={2008},
  doi={10.1103/PhysRevA.78.012331},
  publisher={APS}
}

@article{audenaert2007discriminating,
  title={Discriminating states: The quantum Chernoff bound},
  author={Audenaert, Koenraad MR and Calsamiglia, John and Munoz-Tapia, Ram{\'o}n and Bagan, Emilio and Masanes, Ll and Acin, Antonio and Verstraete, Frank},
  journal={Physical review letters},
  volume={98},
  number={16},
  pages={160501},
  year={2007},
  doi={10.1103/PhysRevLett.98.160501},
  publisher={APS}
}

@article{boutin2017resonator,
  title = {Resonator reset in circuit QED by optimal control for large open quantum systems},
  author = {Boutin, Samuel and Andersen, Christian Kraglund and Venkatraman, Jayameenakshi and Ferris, Andrew J. and Blais, Alexandre},
  journal = {Physical Review A},
  volume = {96},
  issue = {4},
  pages = {042315},
  year = {2017},
  doi = {10.1103/PhysRevA.96.042315}
}

@article{rosenthal2021efficient,
  title={Efficient and low-backaction quantum measurement using a chip-scale detector},
  author={Rosenthal, EI and Schneider, CMF and Malnou, M and Zhao, Z and Leditzky, F and Chapman, BJ and Wustmann, W and Ma, X and Palken, DA and Zanner, MF and others},
  journal={Physical Review Letters},
  volume={126},
  number={9},
  pages={090503},
  year={2021},
  doi={10.1103/PhysRevLett.126.090503},
  publisher={APS}
}

@article{sank2016measurement,
  title={Measurement-induced state transitions in a superconducting qubit: Beyond the rotating wave approximation},
  author={Sank, Daniel and Chen, Zijun and Khezri, Mostafa and Kelly, J and Barends, R and Campbell, B and Chen, Y and Chiaro, B and Dunsworth, A and Fowler, A and others},
  journal={Physical review letters},
  volume={117},
  number={19},
  pages={190503},
  year={2016},
  doi={10.1103/PhysRevLett.117.190503},
  publisher={APS}
}

@article{barends2013coherent,
  title={Coherent Josephson qubit suitable for scalable quantum integrated circuits},
  author={Barends, Rami and Kelly, Julian and Megrant, Anthony and Sank, Daniel and Jeffrey, Evan and Chen, Yu and Yin, Yi and Chiaro, Ben and Mutus, Josh and Neill, Charles and others},
  journal={Physical review letters},
  volume={111},
  number={8},
  pages={080502},
  year={2013},
  doi={10.1103/PhysRevLett.111.080502},
  publisher={APS}
}

@article{riste2012feedback,
  title={Feedback control of a solid-state qubit using high-fidelity projective measurement},
  author={Rist{\`e}, D and Bultink, CC and Lehnert, Konrad W and DiCarlo, L},
  journal={Physical review letters},
  volume={109},
  number={24},
  pages={240502},
  year={2012},
  doi={10.1103/PhysRevLett.109.240502},
  publisher={APS}
}

@article{blais2021circuit,
  title = {Circuit quantum electrodynamics},
  author = {Blais, Alexandre and Grimsmo, Arne L. and Girvin, S. M. and Wallraff, Andreas},
  journal = {Reviews of Modern Physics},
  volume = {93},
  number = {2},
  pages = {025005},
  year = {2021},
  doi = {10.1103/RevModPhys.93.025005}
}

@article{castellanos2008amplification,
  title = {Amplification and squeezing of quantum noise with a tunable Josephson metamaterial},
  author = {Castellanos-Beltran, M. A. and Irwin, K. D. and Hilton, G. C. and Vale, L. R. and Lehnert, K. W.},
  journal = {Nature Physics},
  volume = {4},
  pages = {929--931},
  year = {2008},
  doi = {10.1038/nphys1090}
}

@article{macklin2015near,
  title = {A near–quantum-limited Josephson traveling-wave parametric amplifier},
  author = {Macklin, C. and O’Brien, K. and Hover, D. and Schwartz, M. and Bolkhovsky, V. and Zhang, X. and Oliver, W. D. and Siddiqi, I.},
  journal = {Science},
  volume = {350},
  number = {6258},
  pages = {307--310},
  year = {2015},
  doi = {10.1126/science.aaa8525}
}

@article{heinsoo2018rapid,
  title = {Rapid high-fidelity multiplexed readout of superconducting qubits},
author = {Heinsoo, Johannes and Andersen, Christian Kraglund and Remm, Ants and Krinner, Sebastian and Walter, Theodore and Salath{\'e}, Yves and Gasparinetti, Simone and Besse, Jean-Claude and Poto{\v{c}}nik, Anton and Wallraff, Andreas and Eichler, Christopher},
  journal = {Physical Review Applied},
  volume = {10},
  number = {3},
  pages = {034040},
  year = {2018},
  doi = {10.1103/PhysRevApplied.10.034040}
}

@article{nussbaum2009asymptotic,
  title = {The Chernoff lower bound for symmetric quantum hypothesis testing},
  author = {Nussbaum, Michael and Szko{\l}a, Arleta},
  journal = {Annals of Statistics},
  volume = {37},
  number = {2},
  pages = {1040--1057},
  year = {2009},
  doi = {10.1214/08-AOS593}
}

@article{fowler2012surface,
  title = {Surface codes: Towards practical large-scale quantum computation},
  author = {Fowler, Austin G. and Mariantoni, Matteo and Martinis, John M. and Cleland, Andrew N.},
  journal = {Physical Review A},
  volume = {86},
  number = {3},
  pages = {032324},
  year = {2012},
  doi = {10.1103/PhysRevA.86.032324}
}

@article{terhal2015quantum,
  title = {Quantum error correction for quantum memories},
  author = {Terhal, Barbara M.},
  journal = {Reviews of Modern Physics},
  volume = {87},
  number = {2},
  pages = {307--346},
  year = {2015},
  doi = {10.1103/RevModPhys.87.307}
}

@article{Bengtsson2024,
  title = {Model-Based Optimization of Superconducting Qubit Readout},
  author = {Andreas Bengtsson and Alex Opremcak and Mostafa Khezri and Daniel Sank and Alexandre Bourassa and Kevin J. Satzinger and Sabrina Hong and Catherine Erickson and Brian J. Lester and Kevin C. Miao and Alexander N. Korotkov and Julian Kelly and Zijun Chen and Paul V. Klimov},
  journal = {Physical Review Letters},
  volume = {132},
  number = {10},
  pages = {100603},
  year = {2024},
  doi = {10.1103/PhysRevLett.132.100603}
}

@article{Sunada2024,
  title = {Photon-Noise-Tolerant Dispersive Readout of a Superconducting Qubit using a Nonlinear Purcell Filter},
  author = {Y. Sunada and Y. Kono and Shiyu Wang and Z. Ilves and T. Miyamura and J. Matsuura and S. Tamate and Y. Nakamura},
  journal = {PRX Quantum},
  volume = {5},
  number = {1},
  pages = {010307},
  year = {2024},
  doi = {10.1103/PRXQuantum.5.010307}
}

@article{Spring2025,
  title = {Fast Multiplexed Superconducting-Qubit Readout with Intrinsic Purcell Filtering Using a Multiconductor Transmission Line},
  author = {Peter A. Spring and Luka Milanovic and Yoshiki Sunada and Shiyu Wang and Arjan F. van Loo and Shuhei Tamate and Yasunobu Nakamura},
  journal = {PRX Quantum},
  volume = {6},
  number = {2},
  pages = {020345},
  year = {2025},
  doi = {10.1103/PRXQuantum.6.020345}
}

@article{bultink2016active,
  title={Active resonator reset in the nonlinear dispersive regime of circuit QED},
  author={Bultink, Cornelis Christiaan and Rol, MA and O’Brien, TE and Fu, Xiang and Dikken, BCS and Dickel, Christian and Vermeulen, RFL and De Sterke, JC and Bruno, Alessandro and Schouten, RN and others},
  journal={Physical Review Applied},
  volume={6},
  number={3},
  pages={034008},
  year={2016},
  doi={10.1103/PhysRevApplied.6.034008},
  publisher={APS}
  }

@article{Acharya2025,
  title = {Quantum error correction below the surface code threshold},
  author = {Acharya, Rajeev and others},
  collaboration = {Google Quantum AI},
  journal = {Nature},
  volume = {638},
  pages = {920--926},
  year = {2025},
  doi = {10.1038/s41586-024-08449-y}
}

@article{Ye2024_Quarton,
  title = {Ultrafast superconducting qubit readout with the quarton coupler},
  author = {Ye, Yufeng and Kline, Jeremy B. and Chen, Sean and Yen, Alec and O'Brien, Kevin P.},
  journal = {Science Advances},
  volume = {10},
  number = {41},
  pages = {eado9094},
  year = {2024},
  doi = {10.1126/sciadv.ado9094}
}

@article{Hazra2025,
  title = {Benchmarking the Readout of a Superconducting Qubit for Repeated Measurements},
  author = {Hazra, S. and Dai, W. and Connolly, T. and Kurilovich, P. D. and Wang, Z. and Frunzio, L. and Devoret, M. H.},
  journal = {Physical Review Letters},
  volume = {134},
  number = {10},
  pages = {100601},
  year = {2025},
  doi = {10.1103/PhysRevLett.134.100601}
}

@article{Swiadek2024,
  title = {Enhancing Dispersive Readout of Superconducting Qubits through Dynamic Control of the Dispersive Shift: Experiment and Theory},
  author = {Swiadek, Fran{\c{c}}ois and Shillito, Ross and Magnard, Paul and others},
  journal = {PRX Quantum},
  volume = {5},
  number = {4},
  pages = {040326},
  year = {2024},
  doi = {10.1103/PRXQuantum.5.040326}
}

@article{Dumas2024,
  title = {Measurement-Induced Transmon Ionization},
  author = {Dumas, M. F. and Groleau-Par\'e, B. and McDonald, A. and Muñoz-Arias, M. H. and Lled\'o, C. and D'Anjou, B. and Blais, A.},
  journal = {Physical Review X},
  volume = {14},
  pages = {041023},
  year = {2024},
  doi = {10.1103/PhysRevX.14.041023}

}

@article{Sah2024,
  title = {Decay-protected superconducting qubit with fast control enabled by integrated on-chip filters},
  author = {Sah, Aashish and Kundu, Suman and Suominen, Heikki and Chen, Qiming and M\"ott\"onen, Mikko},
  journal = {Communications Physics},
  volume = {7},
  pages = {227},
  year = {2024},
  doi = {10.1038/s42005-024-01733-3}
}

@article{nesterov2024measurement,
  title={Measurement-induced state transitions in dispersive qubit-readout schemes},
  author={Nesterov, Konstantin N and Pechenezhskiy, Ivan V},
  journal={Physical Review Applied},
  volume={22},
  number={6},
  pages={064038},
  year={2024},
  doi = {10.1103/PhysRevApplied.22.064038},
  publisher={APS}
}

@article{krinner2022realizing,
  title={Realizing repeated quantum error correction in a distance-three surface code},
  author={Krinner, Sebastian and Lacroix, Nathan and Remm, Ants and Di Paolo, Agustin and Genois, Elie and Leroux, Catherine and Hellings, Christoph and Lazar, Stefania and Swiadek, Francois and Herrmann, Johannes and others},
  journal={Nature},
  volume={605},
  number={7911},
  pages={669--674},
  doi = {10.1038/s41586-022-04566-8},
  year={2022},
  publisher={Nature Publishing Group UK London}
}

@article{Arute2019,
  title = {Quantum supremacy using a programmable superconducting processor},
  author = {Arute, Frank and others},
  journal = {Nature},
  volume = {574},
  pages = {505--510},
  year = {2019},
  doi = {10.1038/s41586-019-1666-5}
}

@article{didier2015fast,
  title={Fast quantum nondemolition readout by parametric modulation of longitudinal qubit-oscillator interaction},
  author={Didier, Nicolas and Bourassa, J{\'e}r{\^o}me and Blais, Alexandre},
  journal={Physical review letters},
  volume={115},
  number={20},
  pages={203601},
  year={2015},
  doi={10.1103/PhysRevLett.115.203601},
  publisher={APS}
}

@article{opremcak2021high,
  title={High-fidelity measurement of a superconducting qubit using an on-chip microwave photon counter},
  author={Opremcak, Alexander and Liu, CH and Wilen, C and Okubo, K and Christensen, BG and Sank, D and White, TC and Vainsencher, A and Giustina, M and Megrant, A and others},
  journal={Physical Review X},
  volume={11},
  number={1},
  pages={011027},
  year={2021},
  doi={10.1103/PhysRevX.11.011027},
  publisher={APS}
}

@article{Magnard2018fast,
  title = {Fast and Unconditional All-Microwave Reset of a Superconducting Qubit},
  author = {Magnard, P. and Kurpiers, P. and Royer, B. and Walter, T. and Besse, J.-C. and Gasparinetti, S. and Pechal, M. and Heinsoo, J. and Storz, S. and Blais, A. and Wallraff, A.},
  journal = {Phys. Rev. Lett.},
  volume = {121},
  issue = {6},
  pages = {060502},
  numpages = {6},
  year = {2018},
  month = {Aug},
  publisher = {American Physical Society},
  doi = {10.1103/PhysRevLett.121.060502},
  url = {https://link.aps.org/doi/10.1103/PhysRevLett.121.060502}
}

@article{bahadur1960deviations,
  author  = {Bahadur, R. R. and Ranga Rao, R.},
  title   = {On Deviations of the Sample Mean},
  journal = {The Annals of Mathematical Statistics},
  volume  = {31},
  number  = {4},
  pages   = {1015--1027},
  year    = {1960},
  doi     = {10.1214/aoms/1177705674}
}

@article{efron1968large,
  author  = {Efron, Bradley and Truax, Donald},
  title   = {Large Deviations Theory in Exponential Families},
  journal = {The Annals of Mathematical Statistics},
  volume  = {39},
  number  = {5},
  pages   = {1402--1424},
  year    = {1968},
  doi     = {10.1214/aoms/1177698121}
}

@article{panadero2023photon,
  title   = {Photon-emission statistics for single nitrogen-vacancy centers},
  author  = {Panadero, I. and Espin\'os, H. and Tsunaki, L. and Volkova, K. and
             Tobalina, A. and Casanova, J. and Acedo, P. and Naydenov, B. and
             Puebla, R. and Torrontegui, E.},
  journal = {Phys. Rev. Applied},
  volume  = {22},
  pages   = {014035},
  year    = {2024},
  doi     = {10.1103/PhysRevApplied.22.014035}
}
\end{document}